# Comparison of Analytical and Numerical Resolution Functions in Sputter Depth Profiling

S. Hofmann[(a)], Y. Liu[(b)], J.Y. Wang[(b)], J. Kovac[(c)]

[(a)]Max Planck Institute for Intelligent Systems (formerly MPI for Metals Research), Heisenbergstrasse 3, D-70569 Stuttgart, Germany
[(b)]Department of Physics, Shantou University, 243 Daxue Road, Shantou, 515063 Guangdong, China
[(c)]Department of Surface Engineering and Optoelectronics F4, Jozef Stefan Institute, Jamova 39, 1000 Ljubljana, Slovenia

**Abstract.**
Quantification of sputter depth profiles is frequently done by fitting the convolution integral over concentration and depth resolution function. For a thin delta layer, there exist analytical solutions. The analytical depth resolution functions of two popular approaches, that of the MRI model and that of Dowsett and coworkers are compared. It is concluded that the analytical depth resolution function of the MRI model gives the correct location of a buried delta layer with respect to the measured profile, and a clear description of the information depth in AES, XPS and SIMS. Both analytical solutions can be extended to larger layer thickness. But they are less flexible with respect to physical parameters which are not constant with concentration or sputtered depth, such as detection sensitivity, atomic mixing, roughness or preferential sputtering. For these cases, numerical solutions have to be used.

## 1. Introduction

Sputter depth profiling is a destructive method to directly obtain an image of the in depth distribution of elemental composition. Usually, the primary result is an elemental intensity versus sputtering time. Quantification with respect to original in depth distribution not only requires conversion of the intensity to concentration and the sputtering time to depth, but also a correction of the inevitable distortions induced by the sputtering process [1].
Several approaches to quantification of depth profiles were elaborated during the past thirty years. Most popular are the response function of Dowsett and coworkers [2-5], and the MRI model by Hofmann and coworkers [6-9]. Both approaches are considered in the following, and their basic features are compared. Their common basis is the convolution integral as e.g. discussed by Sanz [10], Zalm et al. [11,12] and Cumpson [13], where the normalized intensity, $I(z)/I^0$ [1] as a function of the sputtered depth $z$ is given by the convolution between the concentration-depth profile $X(z')$ and the depth resolution function (DRF), $g(z)$, which is determined by the physical mechanisms that cause profile degradation [1,8]

$$\frac{I(z)}{I^0} = \int_{-\infty}^{+\infty} X(z')g(z-z')dz' \qquad (1)$$

with $z'$ the running depth parameter for which the composition is defined.

-----------------------------------------------------------------------------------

*Footnote*: Note that in Eq. (1), for simplification an additional noise term in the convolution integral [12] is ignored. Noise is only indirectly important for the quality of fitting the result with measured data, but is it is important for direct deconvolution methods [12-17].

-----------------------------------------------------------------------------------

A direct deconvolution of Eq. (1) to obtain the in-depth concentration $X(z')$ of the analyzed species from measurements of $I(z)/I^0$ is complicated and yields a large scatter enhanced by insufficient data precision (low signal to noise ratio) [13-17]. Therefore it is customary to assume $X(z')$, and to calculate in a "forward" manner – with a known depth resolution function, $g$(z) – the intensity $I(z)/I^0$ and compare it with the measured $I/I^0(z)$. This procedure is performed repeatedly by trial and error until an optimum fit of both is obtained [8,14,18-20]. Nowadays, this is done by a computational program that varies the $X(z')$ parameters until the minimal value of the average deviation $\varepsilon$ of the calculated from the measured profile is achieved (for definition of $\varepsilon$, see Appendix 6.4). The final input $X(z')$ is the reconstructed, original in- depth distribution of composition. Thus, the key to profile quantification is to find the appropriate depth resolution function (DRF) (or response function). Experimentally, the DRF can be determined either by differentiation of the measured profile of a sharp interface of a layer with sufficient thickness, or directly as the measured profile of an infinitesimal thin layer, a so-called delta layer [19,20] (see also Fig. 10). Theoretical modelling of the DRF is usually based on three parameters. In the MRI model, those are the atomic mixing length $w$, the roughness parameter $\sigma$ (i.e the mean deviation of the Gaussian depth distribution), and the information depth parameter $\lambda$. For delta layers, a simple analytical solution of the DRF can be derived, which directly discloses the principle dependencies on the MRI parameters. This DRF is compared with other response functions and with numerical solutions.

## 2. Results and Discussion

### 2.1 The Analytical Depth Resolution Function of the MRI model

Based on the rigorous ballistic relocation calculations of tracer sputtering profiles by Littmark and Hofer [21] and their simplified interpretation by Liau et al. [22], Hofmann established a model for profile quantification based on three fundamental physical parameters: atomic Mixing, Roughness and Information depth. This so-called MRI model was gradually developed during the years 1988-1994 and reported in several publications [6-8,23,24].

Over several years, a number of successful applications were carried out[18,19,25,26] and subsequent refinements and extensions were introduced (see e.g.[1,9,21,27-29]). Although the MRI model was originally developed to quantitatively evaluate depth profiles obtained by sputtering in combination with AES and XPS, one of its first successful applications was the fairly accurate fitting of a SIMS measurement of an AlAs monolayer depth profile in GaAs [24] (see also Fig 6.). However, a monolayer of A in matrix B is more difficult to determine with AES or XPS than with SIMS, because of the much higher sensitivity and dynamic range of the latter technique, as demonstrated in a direct comparison of both techniques in ref. [30]. Therefore, the MRI model was generally applied to obtain depth profiles of typically several nm thickness [8,18]. Nevertheless, for a small layer thickness, e.g. a monolayer, the MRI model can be successfully applied too, as shown in several examples for SIMS [7,17,24,30-33] and AES depth profiles [19,25,26,30,34]. In general, quantitative results of the MRI model are obtained by numerical solution of the convolution integral (Eq. (1)), where the three partial resolution functions combine to $g(z) = g_w \otimes g_\lambda \otimes g_\sigma$. As already shown in ref. [35], for the special case of X(z') being an ideal delta function with vanishing thickness, an analytical resolution function, $g_{\Delta MRI}$, can be derived with the result $I(z)/I^0 = g_{\Delta MRI}$ given by (for derivation see Appendix 6.1)

$$
\begin{aligned}
g_{\Delta MRI}(z) = & \left\{ \frac{1}{2w}\left[1-\exp\left(-\frac{w}{\lambda}\right)\right]\exp\left[\frac{-z-w}{w}+\frac{\sigma^2}{2w^2}\right]\right\} \\
& \times\left\{1-erf\left[\frac{1}{\sqrt{2}}\left(\frac{-z-w}{\sigma}+\frac{\sigma}{w}\right)\right]\right\} \\
& +\left\{\frac{1}{2\lambda}\exp\left[\frac{z}{\lambda}+\frac{\sigma^2}{2\lambda^2}\right]\right\}\times\left\{1+erf\left[\frac{1}{\sqrt{2}}\left(\frac{-z-w}{\sigma}-\frac{\sigma}{\lambda}\right)\right]\right\}
\end{aligned}
\qquad (2)
$$

for a delta layer located at depth $z(0) = 0$. (In reality, z (0) is at a given depth from the surface and therefore $z$ has to be replaced by $z$-z(0)). The MRI parameters are $w$, the mixing length, $\sigma$, the roughness parameter, and $\lambda$, the information depth parameter. These parameters have a well defined physical meaning: $w$ is the length of the zone of complete atomic mixing (also called "cascade mixing"[7,21] in the collision cascade), $\sigma$ is the standard deviation of the Gaussian depth distribution function which characterizes roughness (or, in general, blurred interfaces [20,36,37]), and the information depth parameter $\lambda$ has to be defined according to the method of chemical analysis. In the MRI model, $\lambda$ has a well defined physical meaning. For AES and XPS, $\lambda$ is given by the effective attenuation length (for zero emission angle). In general, $\lambda$ has to be replaced by the mean electron escape depth (MED, see e.g. Chap. 4 in ref. [1])*)

-------------------------------------------------------------------------------------

*Footnote*: *) Here we adopt the original simple notation in ref. [8] and write $\lambda$ for the mean escape depth $\lambda^e$ which is only correct for $\theta = 0$ ($\lambda^e = \lambda\cos\theta$ with $\theta$ the electron emission angle [1]).

-------------------------------------------------------------------------------------

In SIMS, the situation is quite different. The information depth in SIMS is given by the mean depth of origin of the secondary ions. Assuming that practically all of the detected ions stem from the first atomic layer, the information depth parameter $\lambda$ in the MRI model for SIMS can be set to zero (see Sect 2.3.2 for further discussion).

To demonstrate the behavior of the analytical resolution function of the MRI model for different roughness, Fig. 1a shows a plot of Eq. (2) for $w = \lambda = 1$nm and $\sigma = 0.01$, 0.1, 0.3 and 1.0 nm, and Fig. 1b shows the same with a logarithmic intensity scale. Provided that complete atomic mixing is established within the mixing length *w*, the physical behavior is obvious for vanishing roughness (Although the sputtering induced roughness parameter $\sigma$ can hardly be expected to be less than a monolayer [6,7,19,20], the principal behavior of Eq. (2) is better disclosed with that assumption). The steep rise at $z = z(0) - w$ is caused by the actual onset of complete mixing of the delta layer, with the mixing zone length *w* governing the decay of the signal for $z > z(0) - w$. When the roughness increases, this behavior is smoothed out because of the microscopically different spatial onsets of mixing. For vanishing roughness ($\sigma =$ 0.01nm) and assuming that complete atomic mixing is established within the mixing length *w*, the abrupt step upwards at $z = z(0) - w$ is caused by the actual onset of complete mixing of the delta function, with *w* governing the decay of the signal for $z > z(0) - w$. In practice, this behavior (for $\sigma = 0.01$nm) is never observed because the sputtering induced contribution to the roughness parameter $\sigma$ can hardly be expected to be less than a monolayer [6,7,19,20]. Furthermore, a small but non-zero contribution to $\sigma$ may be caused by non-ideal mixing [36]. It is interesting to note that the maximum of the DRF is shifting with increasing roughness from $z = z(0) - w$ in the direction of $z = z(0)$, until it coincides with the centroid of both exponential functions for *w* and $\lambda$ (see Sect. 2.3.1). Increasing roughness causes a deviation from the exponential function in the beginning (given by the information depth parameter $\lambda$). Note that the DRF is normalized so that the integral is unity; therefore every increase in width decreases the according maximum value.

Figs. 2a,b show the influence of decreasing $\lambda$ and increasing *w* values in linear and half-logarithmic plots of the DRF, respectively. For small $\sigma$ (=0.01 nm) and small $\lambda$ ($\leq 0.3$ nm) we approach the asymmetric triangular profile shape with vertical upslope and shift against the delta layer position as discussed and schematically depicted by Zalm and de Kruif (cf. Fig. 1 in ref.[11]). As in Fig. 1a,b, the abrupt step upward for the curves with $\sigma = 0.01$nm are only shown to demonstrate the operational principle, in contrast to the profiles where $\sigma = 1.0$ nm which can be measured in practice.
The analytical DRF given by Eq. (2) looks elegant and fairly simple, but because of the many advantages of the numerical solution of the MRI model (see Sect. 2.5) which is also easy to handle, applications are usually performed with the latter.
A different analytical depth resolution function for SIMS was given by Dowsett et al. [2-4] which is discussed and compared with Eq. (2) in the following.

## 2.2 . The Analytical Response Function of Dowsett et al.[2-4]

In SIMS, the signal acquired for analysis is quite different from AES and XPS, where the surface left after sputter removal is analyzed [1]. Therefore, for homogenous samples in steady state sputtering, AES or XPS detect a surface composition that is modified by preferential sputtering [1] (e.g. after Shimizu et al. [38]). In SIMS, the ionized fraction of the sputtered particles, which represent the bulk composition, is detected. In the dynamic conditions of delta layer depth profiling, at low concentrations and for small sputtering yield differences between analyte and matrix, the resulting difference in the depth profiles appears to be negligible, as demonstrated in ref.[30]. Because matrix effects in intensity measurements are much larger for SIMS, sharp A/B interfaces are less appropriate for experimental DRF determination. However, the high sensitivity enhances the direct measurement of the DRF on sandwich monolayers, where matrix effects are much reduced. SIMS measurements of these “delta layer” structures were frequently characterized by two exponential functions, one with positive slope (upslope) and the other with negative slope (downslope) [10,11,39]. The “missing link” for a full description of a delta layer profile was an additional partial resolution function of Gaussian shape [11] which was already used in profile calculations with the MRI formalism [6,7,23,24]. The latter was for the first time successfully applied for quantification of a SIMS monolayer profile [7,24](see Fig 6). Convolution of the up-and downslope exponentials with a Gaussian was introduced by Dowsett and coworkers [2-4]. In addition, these authors presented an analytical solution of the convolution integral as a delta layer response function or depth resolution function (DRF). Since then, the complete empirical DRF for SIMS, consisting of two exponentials and one Gaussian, was elaborated further and was successfully applied to numerous SIMS depth profiles of thin layers, notably atomic monolayers [39-45]. In 2003, the delta layer response function for SIMS after Dowsett et al. was accepted as an ISO standard by the technical committee TC 201 on Surface Chemical Analysis [4]. The analytical response function or depth resolution function (DRF) of Dowsett and coworkers [2-4] is described by two joined exponentials (up- and downslope) convolved by a Gaussian with mean deviation σ and therefore called UDS in the following. For a delta function at $z(0) = 0$, the DRF in UDS, $g_{\Delta UDS}$, is given by (for derivation see Appendix 6.2)

$$g_{\Delta UDS}(z) = \frac{Q}{2(\lambda_u + \lambda_d)} \left[ \begin{array}{l} \exp\left( \frac{-z}{\lambda_d} + \frac{\sigma^2}{2\lambda_d^2} \right) \left( 1 + erf\left( \frac{1}{\sqrt{2}} \left( \frac{z}{\sigma} - \frac{\sigma}{\lambda_d} \right) \right) \right) \\ + \exp\left( \frac{z}{\lambda_u} + \frac{\sigma^2}{2\lambda_u^2} \right) \left( 1 - erf\left( \frac{1}{\sqrt{2}} \left( \frac{z}{\sigma} + \frac{\sigma}{\lambda_u} \right) \right) \right) \end{array} \right] \qquad (3)$$

with $Q$ a scaling factor (=1 for a normalized delta layer and sensitivity factor = 1), $\sigma$ the standard deviation of an assumed Gaussian distribution function describing roughness (the same meaning as in Eq. (2)), $\lambda_d$ the characteristic downslope or decay length, and $\lambda_u$ the characteristic upslope or growth length. Equation (3) looks somewhat similar to Eq. (2), but there are important differences which will be briefly discussed in the following.

It is obvious that the downslope characteristic length can be associated with the mixing length parameter ($\lambda_d = w$) [1,11,12,43,44], and the parameter $\sigma$ has the same meaning as in the MRI model. The physical meaning of the empirical upslope parameter $\lambda_u$ is still not clear [43-48] and will be discussed below in Sect. 2.3.2. Fig. 3a shows the result of Eq. (3) (normalized to 1, i.e. $Q = 1$), with the same equivalent parameters as in Fig. 1a, and Fig. 3b is the equivalent of Fig. 1b (with logarithmic ordinate). The difference of Figs. 3a,b to Figs 1a,b is obvious. Because the onset of mixing is ignored, the peak of the UDS response function (or DRF) is exactly at the position of the delta layer at $z(0)$, and is not shifted to $z = z(0) - w$ in direction to the surface as in the MRI model. In contrast to Fig. 1b, for increasing roughness, no shift of the maximum is recognized in Fig. 4 (for $\lambda_u = \lambda_d$). The shift of the maximum for $\sigma = 1.0$ nm as a function of $\lambda_d$ for different values of $\lambda_u$ (= 0.3, 0.5 and 1.0 nm) and $\lambda_d$ (=1.0 and 2.0 nm) is shown in Figs. 4a,b, to be compared with Figs. 2a,b. Note that the "unphysical" pointed edge at the maximum for vanishing roughness ($\sigma$ = 0.01 nm) in Figs. 3a.b and 4a,b is analogous to the steep upward step in Figs. 1a.b and 2a,b and the respective discussion applies here too.

### 2.3. Comparison of the Analytical Delta Layer Depth Resolution Functions of the MRI Model and of the UDS Solution

Whereas the mixing length parameter $w$(MRI) is seen to be equivalent to the downslope characteristic length $\lambda_d$(UDS), two major differences are obvious from Figs. 2a,b and 4a,b: (1) The profile shift and (2) the different effects of $\lambda$(MRI) and $\lambda_u$(UDS).

#### 2.3.1 Shift of the measured profile from the original delta layer position

A direct comparison of both the MRI and the UDS delta layer analytical resolution functions is given in Fig. 5a, which in addition shows the influence of a variation of the upslope length $\lambda_u$(UDS) and the information depth parameter $\lambda$(MRI).

Only when $\lambda_u$ is not equal to $\lambda_d$, there is a shift of the maximum in the direction of the centroid of both exponential functions in UDS, hence there is no shift for $\lambda_u = \lambda_d$ (cf. Eq. (5) below). In contrast, in the MRI resolution function there is such a shift even for $\lambda = w$, (see indication in Fig. 1a), since the asymmetry of the profile with respect to the delta layer location is already present for $\sigma \rightarrow 0$. For higher roughness, the original difference in the profile shape decreases. Fig. 5b shows the position of the maximum of the DRF, $z_{max}$, with $\sigma$=1nm, as a function of $w$ for different values of $\lambda$ (= 0.1, 1.0, and 3.0 nm) for MRI, and for UDS with $\sigma$ = 1 nm, as a function of $\lambda_d$ for different values of $\lambda_u$ (= 0.1, 1.0, and 3.0 nm). Fig. 5c depicts $z_{max}$, with $\lambda$ (or $\lambda_u$) = 1nm, as a function of $\sigma$ for different values of $w$ (or $\lambda_d$) for MRI and UDS. The numerical results (not shown here) disclose that the shift of the maximum varies with increasing roughness from the original position at $\sigma \rightarrow 0$ towards the centroid of the respective DRF *) in both MRI and UDS. For UDS, the centroid $\langle z \rangle_{\Delta UDS}$ is given by [36]

$$\langle z \rangle_{\Delta UDS} = \lambda_d - \lambda_u \qquad (4)$$

For MRI, the centroid position $\langle z \rangle_{\Delta MRI}$ is different from Eq. (4) because of the more complicated interaction between $w$ and $\lambda$ and is given by

$$\langle z \rangle_{\Delta MRI} = -\exp(-\frac{w}{\lambda}) \cdot (w + \lambda) \qquad (5)$$

For example, Fig. 5c shows that for MRI with $w = \lambda$ = 1nm at $\sigma \rightarrow 0$ the value of $z_{max}$ = -1.0 nm, and for $\sigma$ = 3nm the value is $z_{max}$ = -0.737nm, in agreement with Eq. (5) ($\langle z \rangle_{\Delta MRI}$ = -0.736 nm). When the DRF is governed by $w$ (i.e. ($\lambda \rightarrow 0$ and $\sigma \rightarrow 0$), it is obvious from the concept of complete mixing that the maximum is shifted in direction of the surface by an amount of $w$ from the delta layer which is at the centroid $\langle z \rangle_{\Delta MRI} = 0$ (Eq. (5) and Fig. 1a).

For UDS, the shift of the maximum is completely different. Because of the formal joint origin of both $\lambda_u$ and $\lambda_d$ at the location of the delta layer, according to Eq. (4) and Fig. 5c, for the corresponding values ($\lambda_d = \lambda_u = 1$ nm), $<z>_{\Delta UDS} = 0$, i.e. it coincides with $z_{max}$ and is independent of $\sigma$ (see Figs, 3a,b). For $\lambda_d > \lambda_u$, the shift of the maximum is in the forward direction (> z(0), see Figs. 4a,b) which is at variance with theoretical predictions (see the comparison MRI-UDS in Fig. 5c and Fig. 9a).

-------------------------------------------------------------------------------------
*) *Footnote*: The centroid is originally given only for the joint exponential functions in UDS [43,44]. Convolution of these by a Gaussian roughness function gives the DRF but that does not change the centroid because the Gaussian is symmetric around z = 0.
-------------------------------------------------------------------------------------

However, it is possible to introduce $\lambda_d = w$ as a shift parameter in the UDS analytical response function for SIMS, and the improved I-UDS function is given by

$$g_{\Delta(I-UDS)} = \frac{1}{2(\lambda_d + \lambda_u)} \cdot \left\{ \begin{aligned} &\exp\left( -\frac{z+\lambda_d}{\lambda_d} + \frac{\sigma^2}{2\lambda_d^2} \right) \cdot \left[ 1 + erf\left( \frac{z+\lambda_d}{\sqrt{2}\sigma} - \frac{\sigma}{\sqrt{2}\lambda_d} \right) \right] \\ &+ \exp\left( \frac{z+\lambda_d}{\lambda_u} + \frac{\sigma^2}{2\lambda_u^2} \right) \cdot \left[ 1 - erf\left( \frac{z+\lambda_d}{\sqrt{2}\sigma} + \frac{\sigma}{\sqrt{2}\lambda_u} \right) \right] \end{aligned} \right\} \quad (6)$$

When we ignore the electron spectroscopic information depth parameter $\lambda$ in the original MRI model and introduce a direct connection of the upslope exponential with the delta layer analogous to UDS, a kind of hybrid MRI (UDS- MRI) follows, with the respective DRF given by (see Appendix 6.1)

$$g_{\Delta(UDS-MRI)} = \frac{1}{2(w + \lambda_u)} \cdot \left\{ \begin{aligned} &\exp\left( -\frac{z+w}{w} + \frac{\sigma^2}{2w^2} \right) \cdot \left[ 1 + erf\left( \frac{z+w}{\sqrt{2}\sigma} - \frac{\sigma}{\sqrt{2}w} \right) \right] \\ &+ \exp\left( \frac{z+w}{\lambda_u} + \frac{\sigma^2}{2\lambda_u^2} \right) \cdot \left[ 1 - erf\left( \frac{z+w}{\sqrt{2}\sigma} + \frac{\sigma}{\sqrt{2}\lambda_u} \right) \right] \end{aligned} \right\} \quad (7)$$

When $w$ is replaced by $\lambda_d$, or vice versa, Eqs. (6) and (7) are identical. If we do not bother about the physical origin of $\lambda_u$ and consider Eqs. (6) and (7) only as a means to fit measured data, we have a function that, in contrast to the original UDS function, takes into account the correct shift of the measured profile from the original delta layer. This fact answers frequently raised questions about the true location of the latter (maximum or centroid of the DRF) [43,44] or that of an interface [46]. The centroid is the right answer [11,46], but not when using the UDS function [46].
(see also Figs.9a and 10)

2.3.2 The upslope length $\lambda_u$ (in UDS) and the information depth parameter $\lambda$ (in MRI)

Of major concern is the difference in $\lambda_u$ (UDS for SIMS) and $\lambda$ (MRI).
Whereas in MRI for AES and XPS the effective attenuation length $\lambda$(MRI) of the emitted electrons is well defined and therefore can be fairly accurately taken from respective databases [1], the situation is more complicated in SIMS.
Because of the similarity of the upslope parameter $\lambda_u$ (UDS) with the partial resolution function of the information depth parameter $\lambda$ in the MRI model, the latter has been frequently taken as one or two atomic monolayers when applied to SIMS depth profiles [19,30]. Indeed, most measurements are fitted with small $\lambda_u$ values of the order of 0.5 – 2 monolayers [12,43-48]. In contrast to the MRI model, the exponential upslope function in the UDS response function is directly folded with a Gaussian function and works like a correction of the latter at lower intensities.

The upslope characteristic length ($\lambda_u$) has been originally [4] and sometimes later [30,47,48] interpreted as the information depth parameter in SIMS. However, a more rigorous consideration shows that this mathematical analogy to $\lambda$ for AES and XPS in the MRI model has no physical meaning in SIMS. It is evident that in the complete mixing model with a fixed value of the mixing length *w*, the onset of the measured intensity is right at the beginning of mixing. Thus, if no other process is present, $\lambda_u$ = 0 should generally apply, as already pointed out by Zalm and de Kruif [11]. By comparison of profiles of B delta layers in Si grown at different temperatures, Dowsett and Chu [39] come to the conclusion that $\lambda_u = 0$ is inherent in SIMS. In a recent paper on the depth of origin of the secondary particles in SIMS, Wittmaack and Mutzke [40] find that in most cases, 98% to 75% of them are from the first monolayer (Note that this applies for atomic ion bombardment and may be different for large cluster ions [47,48]). Thus, there is clear evidence that $\lambda_u$ can be looked upon as an outer parameter, describing an artifact by deviations of instrumental or sample related conditions from the idealized assumptions, useful to interpret experimental results. Since the exponential upslope is frequently observed in SIMS, we have to look for alternative explanations while assuming $\lambda = \lambda_u = 0$.

At first, there is a rather trivial cause for fitting experimental results with an exponential upslope. When the intensity is plotted on a logarithmic scale, the upslope branch of the Gaussian function is of parabolic shape. Away from the maximum, a part of this curve with low curvature can be easily fitted with a straight line describing an exponential function.
An early application of the (numerical) solution of the MRI model with $\lambda = 0$ was the successful fit of a calculated profile to a measured SIMS depth profile of a monolayer of AlAs in GaAs [24]. Fig. 6a presents a re-evaluation of Fig. 6 in ref. [24] (Note that $\lambda_T$ = 3.2 nm per decade of intensity is equal to *w* = 1.3 nm). The measured $Al^+$ ion intensity as a function of sputtered depth reported by Vajo et al. [49] was fairly well fitted with *w* = 1.3 nm, $\sigma$ = 0.8 nm and $\lambda = \lambda_u = 0$ [20]. These original values are only slightly improved in Fig. 6a (with *w* = 1.32 nm, $\sigma$ = 0.75 nm) (Note that MRI and UDS give the same result for $\sigma$ and $w = \lambda_d$ if $\lambda = \lambda_u = 0$, and only the shift by -*w* is different, see Eqs. (6) and (7)). Although we may use Eqs. (6) or (7) with fairly good approximation, we use here Eq. (8) with a monolayer thickness of $z_2$-$z_1$ = 0.28 nm and $z_1$ = 52.10nm, $z_2$ = 52.38nm (see Sect 2.4).

As evident from Fig. 1b, 2b and Fig. 6a, at very low intensities the exponential function finally reaches farther than the Gaussian function and can be clearly distinguished from the latter. Only at lower intensity the straight line indicating the exponential function with $\lambda_u$ is emerging from the parabolic roughness function. By comparison of the derivatives of both functions, Wittmaack [44] derived a critical value for unambiguous detection of $\lambda_u$ which is given by $\lambda_u > \sigma / \sqrt{-2 \ln \eta}$ (for derivation see Appendix 6.3), where $\eta$ denotes the ratio of the critical intensity to the maximum intensity. (Note that this criterion is only a first approximation, because according to Eq. (3) the DRF shape depends on the ratio $\sigma / \lambda_u$).

In Fig. 6a, a deviation of the MRI fit from measurement becomes visible at about 200 counts, that is below about 2% of the maximum intensity. Applying the above criterion for that intensity, $\eta = 0.02$ and $\lambda_u > \sigma/2.8$. With $\sigma = 0.75$ nm, this means that only for $\lambda_u > 0.27$ nm, the latter can be detected without doubt. In accordance with that condition, in Fig. 6b, the MRI and UDS fits are optimized for $\sigma = 0.69$ and 0.68nm, respectively, and for $\lambda$(MRI) = 0.53nm and $\lambda_u$(UDS) = 0.43nm which slightly improves the average deviation $\varepsilon$ (see Appendix 6.4) from 1.10% to 1.05% (see Table 1). Because both $\lambda$ and $\lambda_u$ lack the physical meaning for the SIMS mechanism [11], we may use either MRI or UDS for empirical DRFs. But if we want to understand the physical processes giving rise to $\lambda$ or $\lambda_u$, we have to look for other sources of the exponential upslope.

A not unreasonable cause of the observed upslope may be some kind of contamination effect in the analysis chamber (either by sample preparation or due to redeposition and resputtering [50]). Such an effect was assumed in ref. [19] and was shown to explain the exponential upslope at lower concentrations of a SIMS multilayer profile evaluation. An indication of an additional signal is frequently observed in more or less constant signal intensities at the lower concentration levels before and after the "useful" profile, as seen in Fig. 6a at an intensity level of about 11 counts/s. The result of adding a constant background intensity corresponding to an Al concentration of 0.01 % of a monolayer is shown in Fig. 6c (solid line). Although at the upslope the fit is only slightly improved with respect to Fig. 6a, the regions before the upslope and after the downslope are better reproduced than in Figs 6a and 6b. We conclude that this assumption is principally necessary for each fit and therefore we will keep it for the following improvements.

Until now, we have assumed that the experimental delta layer is exactly a monolayer with infinitely sharp interfaces on both sides (the "layer growers's dream" after P.C. Zalm [12]). However, in reality a monolayer is never atomically sharp because of steps, kinks, dislocations, etc. [12]. Assuming an equivalent of those features as an additional concentration distribution of only 0.3% of a monolayer (ML) stretching over 8 monolayers on each side of the exact delta layer (i.e., a total amount of only 4.8 % of an ML), we get the almost perfect MRI fit shown in Fig. 6d.

During or even after growth, diffusion of Al from the monolayer in the surrounding GaAs can occur, in particular by enhanced "pipe diffusion" [51] through dislocations which may be generated because of stress relief at both interfaces. If we assume an average dislocation density of only about one in 200 atomic distances, this is equivalent to 0.5% of atomic positions of a monolayer with an enhanced diffusivity. Because of the small total change in the monolayer concentration (<10%), we may in good approximation apply the infinite half-space diffusion equation [52] to calculate the analyte distribution shown in Fig. 6e, with a mean diffusion distance of about 2.5 nm (≅ 9 ML) on both sides. These values and the excellent MRI fit (numerical solution) demonstrate that the above assumption is not unreasonable. However, another possible explanation could be an incomplete steep onset of mixing, which generally is included in the Gaussian broadening with the roughness parameter $\sigma$ of the MRI model, as discussed in ref. [36]. When only a few of the delta layer atoms already are transported to the surface before the more abrupt massive mixing sets in, we alternatively may arrive at an explanation similar to Fig. 6e. The latter phenomenon may explain the sometimes found weak dependence of $\lambda_u$ on outer parameters such as the primary ion energy [3].

A comparison of Fig. 6e with Fig. 6d demonstrates that there is always some ambiguousity involved in profile reconstruction [14,18] which can only be minimized by improving profile measurements, that is with both increased experimental depth resolution and improved signal- to- noise figure [1,12,14].

An important phenomenon that is known to influence the shape of the measured profile is preferential sputtering of the species in the delta layer or in the matrix [1,7,9,23]. Introducing the sputtering rate ratio of the delta layer A to the matrix B, $r$(A/B), as a function of the layer material concentration, the expected change of the delta layer (1ML) SIMS profile has already been depicted in Fig. 1 of ref. [7]. For $r>1$, the apparent profile is getting steeper, and for $r<1$, it is getting broader, as shown in Fig. 7. Here, the numerical MRI calculation hast to be applied because the sputtering rate is slightly changing throughout the profile. Since the variation of the analyte concentration is rather small (< 10%), the sputtering time scale is approximately proportional to the depth scale [6]. Fig. 7 may explain the frequently reported different decay length values for different analytes in the same matrix [2].

In summary, there are a number of rather plausible explanations for the experimentally observed upslope exponential function which can be explained by non-exponential functions caused by a variety of effects. This fact explains the somehow “mysterious” role of $\lambda_u$ [44] because of its complex origin, and supports the physically reasonable assumption that the formally corresponding information depth parameter ($\lambda = \lambda_u$) is zero.

### 2.4. Beyond Delta Layers: Analytical and Numerical Solutions of the Convolution Integral for Thick Layers

#### 2.4.1 *Analytical Solution for Thick Layers*

As already proposed by Zalm [12] and later by Gautier et al.[16], including a term for layer thickness appears to be possible in the analytical DRF. For example, in the MRI model we can introduce a layer thickness $d = z_2 - z_1$, where $z_1$ denotes the beginning and $z_2$ the end of the layer. Solving the convolution integral for different parts of the sample, we obtain for the normalized intensity of a thick layer in the MRI model (see Appendix 6.5)

$$
\begin{aligned}
(I(z)/I^0)_{d-MRI} &= \frac{1}{2}\cdot\left[erf\left(\frac{z+w-z_1}{\sqrt{2}\sigma}\right)-erf\left(\frac{z+w-z_2}{\sqrt{2}\sigma}\right)\right] \\
&+\frac{1}{2}\cdot\exp\left(\frac{z}{\lambda}+\frac{\sigma^2}{2\lambda^2}\right) \\
&\times\left\{\exp\left(-\frac{z_1}{\lambda}\right)\cdot\left[1-erf\left(\frac{z+w-z_1}{\sqrt{2}\sigma}+\frac{\sigma}{\sqrt{2}\lambda}\right)\right]-\exp\left(-\frac{z_2}{\lambda}\right)\cdot\left[1-erf\left(\frac{z+w-z_2}{\sqrt{2}\sigma}+\frac{\sigma}{\sqrt{2}\lambda}\right)\right]\right\} \\
&+\frac{1}{2}\cdot\exp\left(-\frac{z+w}{w}+\frac{\sigma^2}{2w^2}\right)\cdot\left[1-\exp\left(-\frac{w}{\lambda}\right)\right] \\
&\times\left\{\exp\left(\frac{z_2}{w}\right)\cdot\left[1+erf\left(\frac{z+w-z_2}{\sqrt{2}\sigma}-\frac{\sigma}{\sqrt{2}w}\right)\right]-\exp\left(\frac{z_1}{w}\right)\cdot\left[1+erf\left(\frac{z+w-z_1}{\sqrt{2}\sigma}-\frac{\sigma}{\sqrt{2}w}\right)\right]\right\}
\end{aligned}
\tag{8}
$$

The respective UDS solution for a thick layer is given by [16]

$$
\begin{aligned}
(I(z)/I^0)_{d-UDS} &= \frac{1}{2}\cdot\left[erf\left(\frac{z-z_1}{\sqrt{2}\sigma}\right)-erf\left(\frac{z-z_2}{\sqrt{2}\sigma}\right)\right] \\
&+\frac{\lambda_u}{2(\lambda_u+\lambda_d)}\cdot\exp\left(\frac{z}{\lambda_u}+\frac{\sigma^2}{2\lambda_u^2}\right) \\
&\times\left\{\exp\left(-\frac{z_1}{\lambda_u}\right)\cdot\left[1-erf\left(\frac{z-z_1}{\sqrt{2}\sigma}+\frac{\sigma}{\sqrt{2}\lambda_u}\right)\right]-\exp\left(-\frac{z_2}{\lambda_u}\right)\cdot\left[1-erf\left(\frac{z-z_2}{\sqrt{2}\sigma}+\frac{\sigma}{\sqrt{2}\lambda_u}\right)\right]\right\} \\
&+\frac{\lambda_d}{2(\lambda_u+\lambda_d)}\cdot\exp\left(-\frac{z}{\lambda_d}+\frac{\sigma^2}{2\lambda_d^2}\right) \\
&\times\left\{\exp\left(\frac{z_2}{\lambda_d}\right)\cdot\left[1+erf\left(\frac{z-z_2}{\sqrt{2}\sigma}-\frac{\sigma}{\sqrt{2}\lambda_d}\right)\right]-\exp\left(\frac{z_1}{\lambda_d}\right)\cdot\left[1+erf\left(\frac{z-z_1}{\sqrt{2}\sigma}-\frac{\sigma}{\sqrt{2}\lambda_d}\right)\right]\right\}
\end{aligned}
\tag{9}
$$

Equations (8) and (9) are for analyte concentration (= mole fraction) and sensitivity factor = 1. For other values, respective scaling factors have to be introduced.

Equation (8) shows that the effect of the information depth in AES and XPS on the intensity can be fully included in the analytical solution, in contrast to an earlier prediction [37] where this was thought to be impossible. As in the case of a delta layer, the advantage of Eq. (8) over Eq. (9) is the correct location of the measured profile with respect to the interfaces at $z_1$ and $z_2$.

With Eqs. (8) or (9), the old question whether a monolayer is the equivalent of an ideal delta layer with infinitesimal small thickness (d→0) can easily be studied. Particularly in SIMS, the simple analytical solution of the ideal delta layer is usually applied for monolayers (see Eqs. (2),(3)). In reality, however, the thinnest layer is an atomic monolayer, with a thickness of 0.25 ± 0.05 nm in most semiconductors and metals. If we assume a DRF of lower limit, for example about one atomic monolayer for each of the parameters information depth, roughness and mixing, the resulting

FWHM of the profile for $z_2 - z_1 = 0$ is about 3.3 monolayers or ca. 1 nm for a delta layer [19,20]. As depicted in Fig. 8a (similar to Fig. 8 in ref.[53]), the FWHM of the measured layer profile after Eq. (8) increases slightly with increasing layer thickness until it becomes identical to the latter for a thickness above about 8 monolayers. It is interesting to note that even between $d = 0$ (ideal delta layer) and $d$= 1ML there already is a slight difference (ca. 2.7%) and the error in the FWHM of the profile when applying Eq. (2) to 2 ML is about 10%. In Fig. 8b, a diagram similar to Fig. 8a is shown for the case of SIMS with $\lambda_u = \lambda = 0$, $\lambda_d = w = \sigma =$ 1ML, which yields an increased deviation. In any case, for higher values of the DRF parameters the deviation between the ideal delta layer and a monolayer is reduced.

In order to emphasize the physical relevance of the MRI parameters and the power of the analytical solution of Eq. (1) for different layer thickness, Eq. (8) is applied to depth profiles obtained by much more sophisticated approaches. Figure 9 shows tracer profiles of two sandwich layers (of isotopes) of 1 nm (Fig. 9a) and 30 nm (Fig. 9b) thickness, calculated by Littmark and Hofer [21] for 5 keV $Ar^+$ ions at normal incidence. The authors used Sigmund's theory for ballistic relocation, including diffusion approximation in the mixing zone, sputter removal of the isotope layer and SIMS type analysis. The much simpler calculation of $(I(z)/I^0)_{d\text{-MRI}}$ by Eq. (8) yields an acceptable fit of the results shown in ref. [21]. The profile shift with respect to the original in depth distribution of composition is correctly reproduced, using the ideal SIMS approximation with $\lambda \cong 0$, and w = 6.49 nm is in reasonable agreement with the damage distribution of about 6 nm calculated by Littmark and Hofer [21]. Of course, with adequate parameters ($\lambda_u \cong 0$), Eq. (9) for UDS gives the same fit of the profile shape, but with a deviating position of the original layer at z(0) ≅ 23.6 nm, in contrast to ref. [21] where z(0) ≅ 30.5 nm.

The analytical solution of the MRI model for thick layers (Eq. (8)) is extremely useful when the apparent location of interfaces in the measured profile of thin layers with increasing thickness have to be compared with the true layer thickness to find the change of the sputtering rate with depth. Such a case was reported by Seah et al. [46] for SIMS depth profiles of $SiO_2$ layers with different thickness obtained by bombardment with 600 eV $Cs^+$ ions. These authors employed the UDS response function (Eq.(3)) which could not yield the physically relevant centroid positions (see Sect. 2.3.1), and they finally assumed the location of the interface at 50% of the plateau intensity. In contrast, the MRI model gives the correct DRF centroid position and discloses its different location with respect to the 50% interface position (and additionally with respect to the DRF or delta layer profile peak position). This is shown in Fig. 10 for $\lambda = 0$, $w$ = 1nm and different $\sigma$ values. The shift of the 50% interface position decreases with increasing $\sigma$ from $w$*(1-ln2) for $\sigma = 0$ to zero for predominant $\sigma$ ($\sigma$ >>w), when it coincides with the centroid of the DRF in MRI.

In summary, analytical DRFs can be applied to the convolution integral of

1) Delta layers (Eqs. (2),(3))
2) Layers with any finite thickness and constant analyte concentration (Eqs. (8), (9))
3) Multilayers of type 2)
4) Layers with concentration gradients can be approximated by an array of sublayers with stepwise changes of concentration and different thickness (Eqs. (8), (9))

Two of the three MRI parameters ($\lambda$, w) have to remain constant throughout the profile, as well as additional parameters such as a concentration dependent sputtering rate ratio (preferential sputtering). However, for the roughness parameter $\sigma$, a depth-dependent term can be introduced into Eqs. (2) and (8) (see explanation in App. 6.5).

Of course, numerical solutions for profile reconstruction are always possible with practically no restriction.

2.4.2 *Numerical Solutions for Thick Layers*

Analytical resolution functions are particularly user-friendly because for the straight forward calculation no computer program (as for numerical solutions) is necessary. But the analytical solution of the convolution integral cannot be performed with an analytical DRF in case of any change of two of the three parameters ($w$, $\lambda$) during profiling, and of the sputtering time/depth and the intensity/concentration relations with the sputtered depth. (For the parameter $\sigma$, the situation is less restrictive (see Appendix A6.5)) In the above mentioned cases, a numerical solution of Eq. (1) is required, of which a few examples were shown in the following references:

1) Depth or concentration dependence of any of the three physically well defined parameters $\lambda$ [34], $w$ [9], and $\sigma$ [9,28]
2) Nonlinear time/depth dependence (preferential sputtering) [1,9]
3) Nonlinear intensity/concentration dependence (e.g. concentration dependent sensitivity factor or secondary cluster ions [30]).
4) Electron backscattering in AES sputter depth profiling [27].

## 3. Conclusion

The convolution integral describing sputter depth profiling in SIMS, AES, XPS and ISS can be solved analytically for infinitesimal thin delta layers as well as for layers with finite thickness. The user - friendly analytical depth resolution function of the MRI model for delta layers is presented and compared with that after Dowsett et al. (UDS) in order to explain the basic differences. Although both approaches are strong simplifications of the sputtering process, it is evident that the MRI model is better representing the physics of the ion bombardment induced processes (like the onset of atomic mixing), and the different analysis methods (AES, XPS or SIMS and ISS profiles). Thus, the MRI model gives a correct value of the shift of the measured profile from the original position of the delta layer or the interface. Since SIMS inherently produces no upslope profile, the according characteristic length is basically zero in both approaches ($\lambda_u$(UDS) = $\lambda$(MRI) = 0). The frequently determined exponential function for the experimental upslope has to be ascribed to additional experimental conditions or parameters such as contamination by resputtering and redeposition, deviation of the in-depth-distribution from ideal sharp interfaces, preferential sputtering etc. In reality, there is no "infinitesimal thin" delta layer, since the lowest possible layer thickness is one atomic monolayer. Whereas the slight deviation between the ideal delta layer and the monolayer can usually be ignored, for thicker layers the analytical delta layer resolution function has to be extended to explicitly include the layer thickness. Analytical solutions of the convolution integral are restricted to constant mixing and information depth parameters and to a constant sputtering rate. Variations of all those parameters during depth profiling can be included in numerical solutions for profile reconstruction. However, it should be kept in mind that they only can represent a more or less simplified and incomplete image of the complex physical processes in sputter depth profiling.

## 4. Acknowledgement

Helpful discussions with Martin Noah, Max-Planck-Institute for Intelligent Systems, Stuttgart, Germany, and with Holger Hofmann, Hiroshima University, Japan, are gratefully acknowledged. This work is supported by the National Natural Science Foundation of China (Project No. 11274218).

## 6. Appendix

**6.1** Derivation of the analytical DRF for delta layers in the MRI model

In the MRI model, the three partial DRFs (given by the parameters mixing length ($w$), roughness ($\sigma$) and information depth ($\lambda$) (e. g. in AES, XPS) establish the total depth resolution function.

$$\begin{cases} g_w(z) = \frac{1}{w}\exp\left(-\frac{z+w}{w}\right) & z > -w \\ g_\sigma(z) = \frac{1}{\sqrt{2\pi}\sigma}\exp\left(-\frac{z^2}{2\sigma^2}\right) & \\ g_\lambda(z) = \frac{1}{\lambda}\exp\left(\frac{z}{\lambda}\right) & z \le 0 \end{cases} \quad \text{(A1)}$$

Ignoring the interplay between atomic mixing and signal detection, the DRF is convolved directly by Eq. (A1), i.e. $g(z) = g_w \otimes g_\lambda \otimes g_\sigma$. Thus, convolution of $g_w$ with $g_\lambda$ results in:

$$g'_{w\lambda}(z) = g_w(z) \otimes g_\lambda(z) = \begin{cases} \frac{1}{w+\lambda}\exp(\frac{z+w}{\lambda}) & z < -w \\ \frac{1}{w+\lambda}\exp(-\frac{z+w}{w}) & z \ge -w \end{cases} \quad \text{(A2)}$$

However, from a physical point of view, the measured signal from the mixing zone is detected from the onset of atomic mixing at $z=z(0)-w$ (for the analyzed delta layer at $z(0) = 0$). For $z > z(0)-w$, the complete mixing zone is analyzed. (see details in refs.[8, 37]. Thus, the partial DRF when combining $g_w$ and $g_\lambda$ is given by:

$$g_{w\lambda}(z) = \begin{cases} g_\lambda(\mathrm{z}) & z \le -w \\ g_w(z) \cdot \int_z^{z+w} g_\lambda(z-z')dz' & z > -w \end{cases}$$

$$= \begin{cases} \frac{1}{\lambda} \cdot \exp(\frac{z}{\lambda}) & z \le -w \\ \frac{1}{w} \cdot \exp(-\frac{z+w}{w}) \cdot \left[1 - \exp(-\frac{w}{\lambda})\right] & z > -w \end{cases} \quad \text{(A3)}$$

The convolution between the first and the second partial terms of Eq. (A3) with their restricted conditions and a Gaussian function results in:

$$\frac{1}{\lambda} \cdot \exp(\frac{z}{\lambda}) \otimes g_\sigma(z) = \frac{1}{2\lambda}\exp\left(\frac{z}{\lambda} + \frac{\sigma^2}{2\lambda^2}\right)\left[1 + erf\left(-\frac{z+w}{\sqrt{2}\sigma} - \frac{\sigma}{\sqrt{2}\lambda}\right)\right] \quad \text{(A4-1)}$$

and

$$\frac{1}{w} \cdot \exp\left(-\frac{z+w}{w}\right) \cdot \left[1 - \exp(-\frac{w}{\lambda})\right] \otimes g_\sigma(z)$$

$$= \frac{1}{2w}\left[1 - \exp(-\frac{w}{\lambda})\right]\exp\left(-\frac{z+w}{w} + \frac{\sigma^2}{2w^2}\right)\left[1 - erf\left(-\frac{z+w}{\sqrt{2}\sigma} + \frac{\sigma}{\sqrt{2}w}\right)\right] \quad \text{(A4-2)}$$

The sum of Eqs. (A4-1), (A4-2) gives $g_{\Delta MRI} = g_{w\lambda} \otimes g_\sigma$, which is the MRI analytical solution (Eq. (2)).

Similarly, $g_{\Delta(UDS\text{-}MRI)}$ (see Eq. (7)) is derived by convoluting $g'_{w\lambda}$ (Eq. (A2)) with $g_\sigma$.

**6.2** Derivation of the UDS expression for delta layers

Similar to the derivation of the MRI function in App. 6.1, the UDS response function is derived from the convolution two exponential functions representing upslope ($\lambda_u$) and downslope ($\lambda_d$) and a Gaussian function ($\sigma$), given by

$$\begin{cases} g_d(z) = \dfrac{1}{\lambda_d}\exp(-\dfrac{z}{\lambda_d}) \\ g_\sigma(z) = \dfrac{1}{\sqrt{2\pi}\sigma}\exp\left(-\dfrac{z^2}{2\sigma^2}\right) \\ g_u(z) = \dfrac{1}{\lambda_u}\exp(\dfrac{z}{\lambda_u}) \end{cases} \qquad \text{(A5)}$$

The two exponential functions $g_d$ and $g_u$ are convolved with each other giving

$$g_{du}(z) = g_d(z) \otimes g_u(z) = \begin{cases} \dfrac{1}{\lambda_d + \lambda_u}\exp(\dfrac{z}{\lambda_u}) & z < 0 \\ \dfrac{1}{\lambda_d + \lambda_u}\exp(-\dfrac{z}{\lambda_d}) & z \geq 0 \end{cases} \qquad \text{(A6)}$$

The convolution between these two terms described by Eq. (A6) and a Gaussian function, $g_\sigma$ results in:

$$\begin{aligned} &\frac{1}{\lambda_d + \lambda_u}\cdot\exp\left(\frac{z}{\lambda_u}\right)\otimes g_\sigma(z) \\ &= \frac{1}{2(\lambda_d + \lambda_u)}\cdot\exp\left(\frac{z}{\lambda_u} + \frac{\sigma^2}{2\lambda_u^2}\right)\cdot\left\{1 - erf\left[\frac{1}{\sqrt{2}}\left(\frac{z}{\sigma} + \frac{\sigma}{\lambda_u}\right)\right]\right\} \end{aligned} \qquad \text{(A7-1)}$$

and

$$\begin{aligned} &\frac{1}{\lambda_d + \lambda_u}\cdot\exp\left(-\frac{z}{\lambda_d}\right)\otimes g_\sigma(z) \\ &= \frac{1}{2(\lambda_d + \lambda_u)}\cdot\exp\left(-\frac{z}{\lambda_d} + \frac{\sigma^2}{2\lambda_d^2}\right)\cdot\left\{1 + erf\left[\frac{1}{\sqrt{2}}\left(\frac{z}{\sigma} - \frac{\sigma}{\lambda_d}\right)\right]\right\} \end{aligned} \qquad \text{(A7-2)}$$

Eq. (3) is given by Eqs. (A7-1) + (A7-2)

**6.3** Wittmaack criterion [44] for distinction between roughness ($\sigma$) and upslope parameter ($\lambda_u$) in the UDS response function (also applicable to $\sigma$ and $\lambda$ in the MRI model)

The Gaussian function ( $g_\sigma(z) = \dfrac{1}{\sqrt{2\pi}\sigma}\exp\left(-\dfrac{z^2}{2\sigma^2}\right)$ ) for roughness and the exponential function ( $g_u(z) = \dfrac{1}{\lambda_u}\exp\left(\dfrac{z}{\lambda_u}\right)$ ) for upslope are assumed to be independent. Their differentiation in semi-logarithmic coordinates are given as $\frac{d\left[\ln(g_\sigma)\right]}{dz} = -\frac{z}{\sigma^2}$ and $\frac{d\left[\ln(g_u)\right]}{dz} = \frac{1}{\lambda_u}$.

For a safe determination of the upslope exponential function, its slope must be smaller than that of the Gaussian, i.e.
$\frac{d\left[\ln(g_\sigma)\right]}{dz} > \frac{d\left[\ln(g_\lambda)\right]}{dz}$ which results in $-z < \frac{\sigma^2}{\lambda_u}$. If $\lambda_u$ is to be determined safely at a level $\eta$ relative to the peak height of the profile, $-z_\eta < \frac{\sigma^2}{\lambda_u}$ follows and is estimated as the position of $\eta$: $z_\eta = -\sigma\sqrt{-2\ln\eta}$ after
$g_\sigma(z_{1/\eta}) = \frac{1}{\sqrt{2\pi}\sigma}\exp\left(-\frac{z_{1/\eta}^2}{2\sigma^2}\right) = \frac{1}{\eta}\cdot\frac{1}{\sqrt{2\pi}\sigma}$. Thus, Wittmaack's criterion is given as

$$\lambda_u > \frac{\sigma}{\sqrt{-2\ln\eta}} \qquad \text{(A8)}$$

**6.4** Definition of the average deviation between a measured an a fitted profile

In ref. [29], we proposed a convenient quantitative description of the quality of a fit of a measured profile, $(I_A/I^0_A)_{meas}$, to a calculated profile, $(I_A/I^0_A)_{calc}$, by the average deviation between both quantities. This average deviation, $\varepsilon$, is given by the sum of all chi-square deviations for n points along the profiles, divided by the number $n$, normalized to the maximum "measured intensity" $(I_A/I_A^0)_{meas}^{\max}$ ,and given in % of the latter:

$$\varepsilon(\%) = 100 \times \sqrt{\frac{\sum_i^n \left[(I_A/I_A^0)_{calc} - (I_A/I_A^0)_{meas}\right]^2}{n\cdot(I_A/I_A^0)_{meas}^{\max}}} \qquad \text{(A9)}$$

The depth range is usually restricted to a region where the intensity exceeds 0.01% of $(I_A/I_A^0)_{meas}^{\max}$. An acceptable good fit should be in the lower percentage region, above 5% the average deviation is not considered to be acceptable.

**6.5** Derivation of the analytical expression for thick layers in the MRI model

For a thick layer with constant concentration of 1 (mol or 100% of the analyte), whose concentration profile is given by $g_d(z) = 1$ $(z_1<z<z_2)$, where $z_1$ and $z_2$ are the start and the end positions of a layer with thickness $d = z_2 - z_1$, its measured profile after atomic mixing can be deduced analytically by $g_{d\text{-}w\lambda} = g_{w\lambda} \otimes g_d$ and is obtained as

$$
g_{d-w\lambda}(z)=\begin{cases}
\exp(\frac{z}{\lambda})\cdot\left[\exp(-\frac{z_1}{\lambda})-\exp(-\frac{z_2}{\lambda})\right] & z<z_1-w\\
\left[1-\exp(-\frac{w}{\lambda})\right]\cdot\left[1-\exp(\frac{z_1-z-w}{w})\right] & \\
+\exp(-\frac{w}{\lambda})\cdot\left[1-\exp(-\frac{z_2-z-w}{\lambda})\right] & z_1-w<z<z_2-w\\
\left[1-\exp(-\frac{w}{\lambda})\right]\cdot\exp(-\frac{z+w}{w})\cdot\left[\exp(\frac{z_2}{w})-\exp(\frac{z_1}{w})\right] & z>z_2-w
\end{cases}
\qquad \text{(A10)}
$$

From a physical point of view, the above convolution may be understood as the detected intensity is generated from all the delta layers of which constructed the thick layer is built up. The profile described by the three partial terms in Eq. (A10) is then convoluted with a Gaussian function $g_\sigma$ for the roughness effect, and the three resulting expressions are derived as

$$
\begin{cases}
F_1(z)=\frac{1}{2}\cdot\left[\exp\left(-\frac{z_1}{\lambda}\right)-\exp\left(-\frac{z_2}{\lambda}\right)\right]\cdot\exp\left(\frac{z}{\lambda}+\frac{\sigma^2}{2\lambda^2}\right)\\
\times\left[1-erf\left(\frac{z+w-z_1}{\sqrt{2}\sigma}+\frac{\sigma}{\sqrt{2}\lambda}\right)\right]\\
F_2(z)=\frac{1}{2}\cdot\left[erf\left(\frac{z+w-z_1}{\sqrt{2}\sigma}\right)-erf\left(\frac{z+w-z_2}{\sqrt{2}\sigma}\right)\right]\\
-\frac{1}{2}\cdot\exp\left(\frac{z-z_2}{\lambda}+\frac{\sigma^2}{2\lambda^2}\right)\cdot\left[erf\left(\frac{z+w-z_1}{\sqrt{2}\sigma}+\frac{\sigma}{\sqrt{2}\lambda}\right)-erf\left(\frac{z+w-z_2}{\sqrt{2}\sigma}+\frac{\sigma}{\sqrt{2}\lambda}\right)\right]\\
-\frac{1}{2}\cdot\left[1-\exp\left(-\frac{w}{\lambda}\right)\right]\cdot\exp\left(-\frac{z+w-z_1}{w}+\frac{\sigma^2}{2w^2}\right)\\
\times\left[erf\left(\frac{z+w-z_1}{\sqrt{2}\sigma}-\frac{\sigma}{\sqrt{2}w}\right)-erf\left(\frac{z+w-z_2}{\sqrt{2}\sigma}-\frac{\sigma}{\sqrt{2}w}\right)\right]\\
F_3(z)=\frac{1}{2}\cdot\left[1-\exp\left(-\frac{w}{\lambda}\right)\right]\cdot\left[\exp\left(\frac{z_2-w}{w}\right)-\exp\left(\frac{z_1-w}{w}\right)\right]\\
\times\exp\left(-\frac{z}{w}+\frac{\sigma^2}{2w^2}\right)\cdot\left[1+erf\left(\frac{z+w-z_2}{\sqrt{2}\sigma}-\frac{\sigma}{\sqrt{2}w}\right)\right]
\end{cases}
\qquad \text{(A11)}
$$

The final expression of the MRI intensity profile for a thick layer is obtained by the sum of the three partial terms in Eq. (A11):

$$
(I(z)/I^0)_{d-MRI}=F_1(z)+F_2(z)+F_3(z) \quad \text{(A12)}
$$

and given by Eq. (8).

Since the measured profile $(I(z)/I^0)_{d\text{-}MRI}$ (Eq. (A12 ) is obtained by convolution with σ after combination of w with λ, the broadening action on the profile only depends on the value of σ at the corresponding depth . Therefore the variation of $\sigma$ with z can be directly introduced into Eq. (8) as an extension. This is also valid for Eq. (2).

Fig. A6.5-1shows an example for a 10 nm thick layer and the depth profile with the MRI parameters: $w = \lambda = 0.3$nm; $\sigma = 1+0.57\sqrt{z}$ (nm) after Eq. (A12), where the equivalent numerical solution is also shown,

**Figure captions:**

**Fig. 1a:** Analytical depth resolution function of the MRI model (Eq. (2)) for $w = \lambda$ = 1nm and different roughness parameter values, $\sigma$ = 0.01, 0.1, 0.3 and 1nm

**Fig. 1b:** As Fig. 1a, but with logarithmic ordinate.

**Fig. 2a**: Influence of different MRI parameters (shown in the inset) on the analytical DRF. Note the increasing shift of the maximum for higher $\lambda$ and for higher $w$ values.

**Fig. 2b**: As Fig. 2a but with logarithmic ordinate.

**Fig. 3a:** The UDS response function (Eq. (3)) in linear coordinates for $\lambda_d = \lambda_u$ = 1nm and different roughness parameter values, $\sigma$ = 0.01, 0.1, 0.3 and 1nm as in Fig. 1a.

**Fig. 3b**: As Fig. 2a but with logarithmic ordinate.

**Fig. 4a**: Influence of different parameters (shown in the inset, with equivalent values as in Fig. 2a) on the UDS response function.

**Fig. 4b**: As Fig. 4a, but with logarithmic ordinate.

**Fig. 5a**: Direct comparison of MRI and UDS analytical depth resolution functions for $w$ = 1.0 nm and with different values of $\lambda = \lambda_u$ and $\sigma$ shown in the inset.

**Fig. 5b**: Shift of the position of the maximum $z_{max}$ of the measured profiles for $\sigma$ = 1.0 nm, and $\lambda$ (=$\lambda_u$) = 0.1, 1.0 and 3.0 nm as a function of the mixing length $w$ (=$\lambda_d$) for MRI and UDS.

**Fig. 5c**: Comparison of the shift of the maximum for $w$ (=$\lambda_d$) = 1.0 nm, and $\lambda$ (=$\lambda_u$) = 0.1, 1.0 and 3.0 nm, as a function of the roughness parameter $\sigma$ for UDS and MRI (similar to Fig. 5b).

**Fig. 6:** SIMS depth profile of Al for an AlAs monolayer (ML) in GaAs using 1 keV $O_2^+$ with 63° incidence angle. The experimental data are from Vajo et al. [49]. (see also Table 1)
(a) MRI analytical DRF fit for $\lambda$ = 0 (from ref. [24], slightly improved)
(b) as in (a), but with $\lambda$(MRI) and $\lambda_u$(UDS) (shown in the inset) as additional fit parameter.
(c) as in (a), but fit using the numerical MRI model with $\lambda$ = 0 and an assumed constant background signal of 11 counts (corresponding to 0.01% of a ML).
(d) as in (c), but with an additional uniform distribution of 0.3% of a ML stretching over 8 ML on each side
(e) as in (c), but with an additional diffusion profile according to $4.5*10^{-3}\{1\text{-erf}[(z\text{-}z_0)/2.5]\}$[52] (with a maximum at 0.45% of a ML) on both sides, simulating a limited pipe diffusion broadening of the ideal monolayer previous to depth profiling.

**Fig. 7**: Measured sputtering depth profile obtained by MRI calculations for different relative sputtering rate ratios *r* (=0.5, 1, 2) of delta layer and matrix. Numerical solutions are obtained for delta layer thickness of $d = z_2 - z_1$ = 1ML, *w* = 10ML, and $\sigma$ = 2.5ML. The original time scale is replaced by a depth scale. Both are nonlinearly related. However, because of the low analyte concentration, the deviation from linearity never exceeds 1%. Adapted from Fig. 1 of ref. [7].

**Fig. 8 (a)**: Comparison of the full width at half maximum (FWHM) of MRI profile calculations for different layer thickness between *d* = 0.3 and *d* = 10 ML after Eq. (8) with the FWHM of the respective ideal delta layer ($d \rightarrow 0$) (Eq. (2)) (solid line), and the MRI parameters $w = \sigma = \lambda$ = 1ML . The dashed line denotes *d* = FWHM valid for larger thickness.
(b) The same as in (a), using the UDS approach Eq. (9) with $\lambda_d = \sigma$ = 1ML and $\lambda_u = 0$.

**Fig. 9**: Transport theory calculations of Littmark and Hofer [21] for tracer sputter depth profiles of Si* in Si, (a) thickness 1nm, (b) thickness 30nm, fitted by MRI profile calculations with the same parameters for (a) and (b) (*w* = 6.94nm, $\sigma$ = 3.08nm and $\lambda$ = 0.01nm), with average deviation for (a) $\varepsilon$ = 2.68% and for (b) $\varepsilon$ = 2.0% (see Appendix 6.4). The centroid for MRI coincides with the layer given in ref, [21] (at 30,5 nm), while that of UDS is 6.94 nm below (at 23.6 nm)

**Fig. 10**: MRI model applied to a sharp interface (use of eq. (8)) showing the shift of the "measured" (apparent) interface from the true interface position at *z* = 0, as compared to a delta layer at *z*=0, for $\lambda = 0$ and $\sigma$ = 0.2, 0.5 and 1.0 nm. Note the shift of the 50% interface.

**Fig. A6.5-1:** Depth profile of a 10 nm thick layer with the MRI parameters: $w = \lambda$ = 0.3nm; $\sigma = 1+0.57\sqrt{z}$ after Eq. (A12), showing the equivalent analytical and numerical solution of the convolution integral.

**Figures:**

(a)

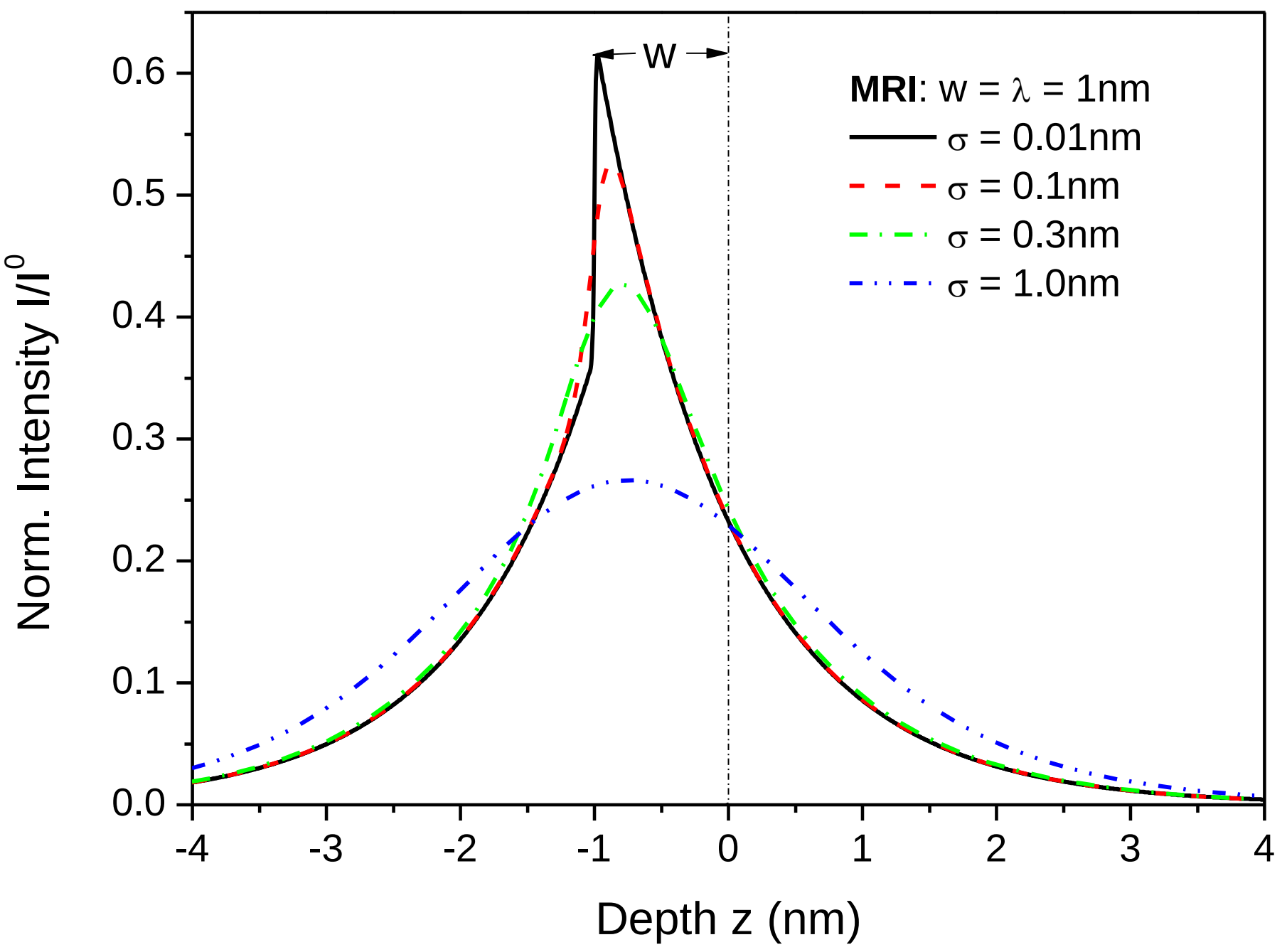

Fig. 1a

(b)

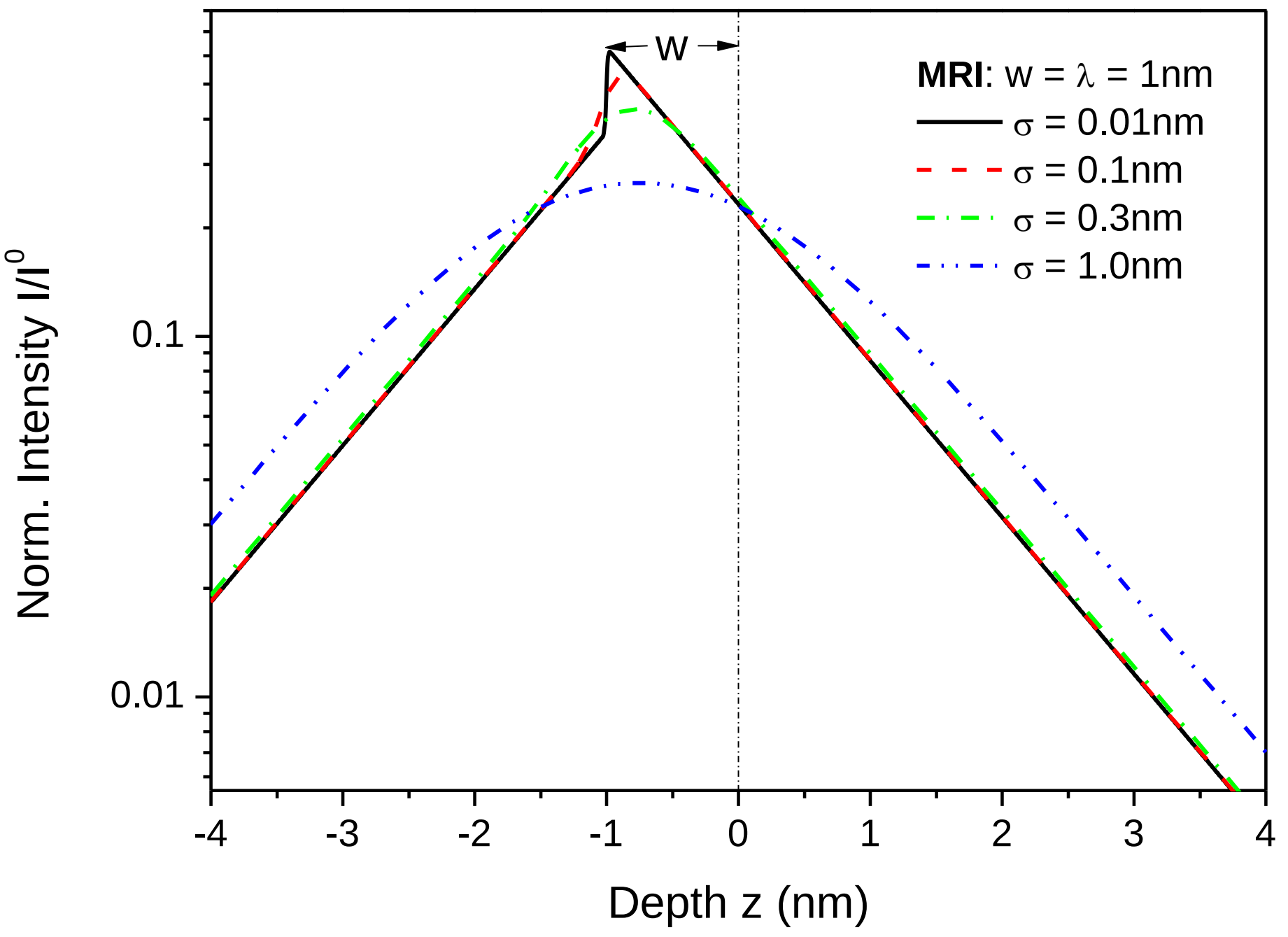

Fig. 1b

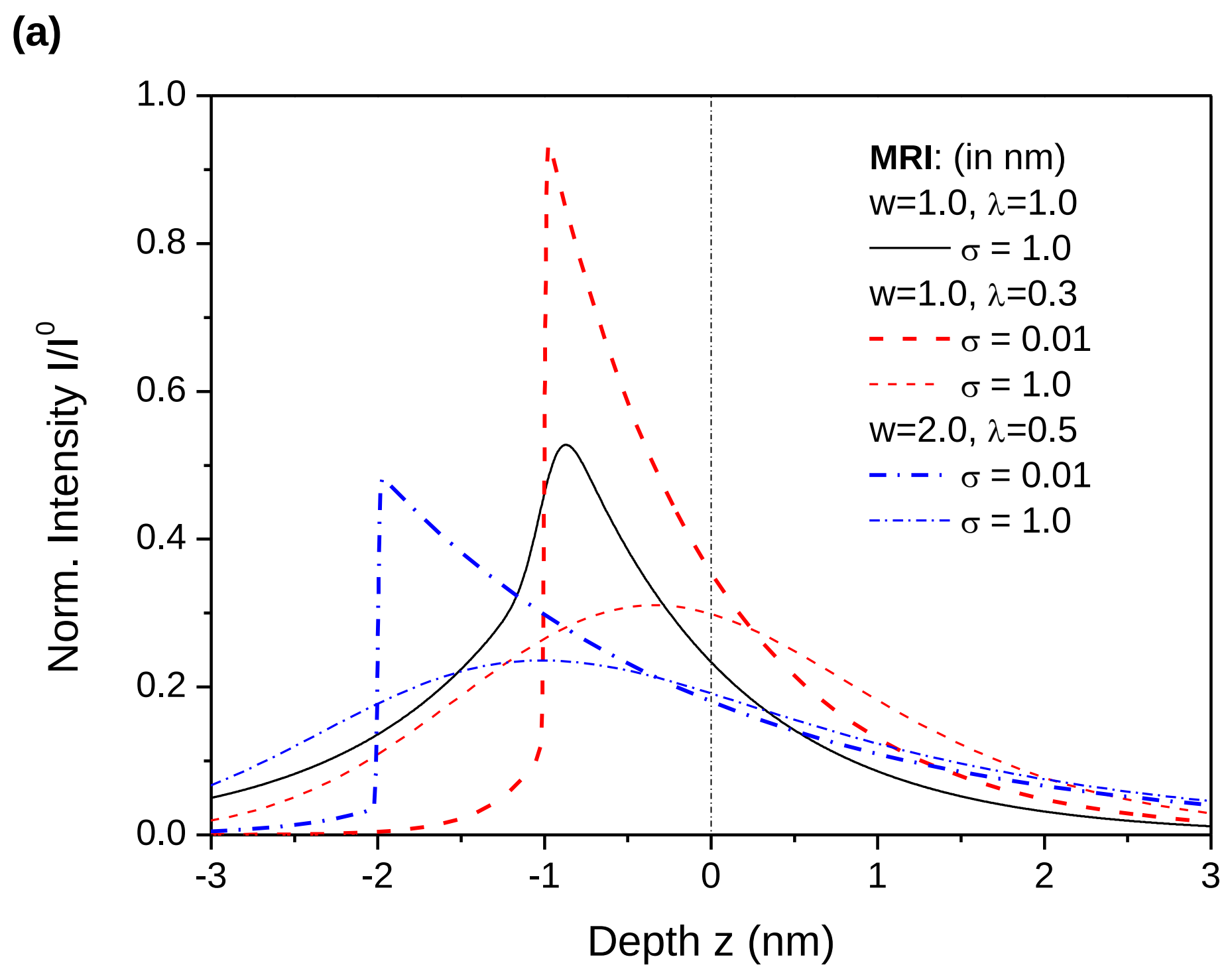

Fig. 2a

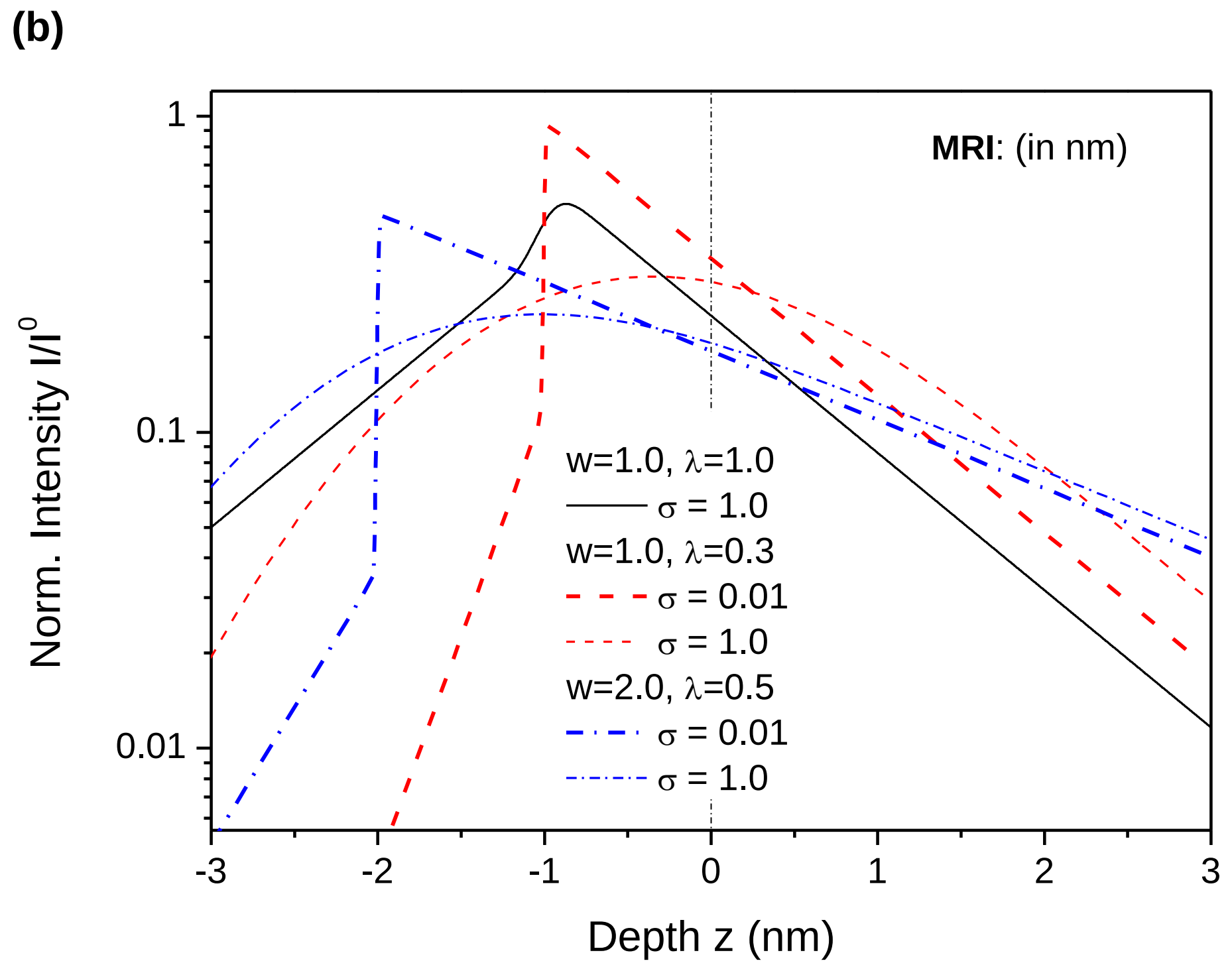

Fig. 2b

(a)

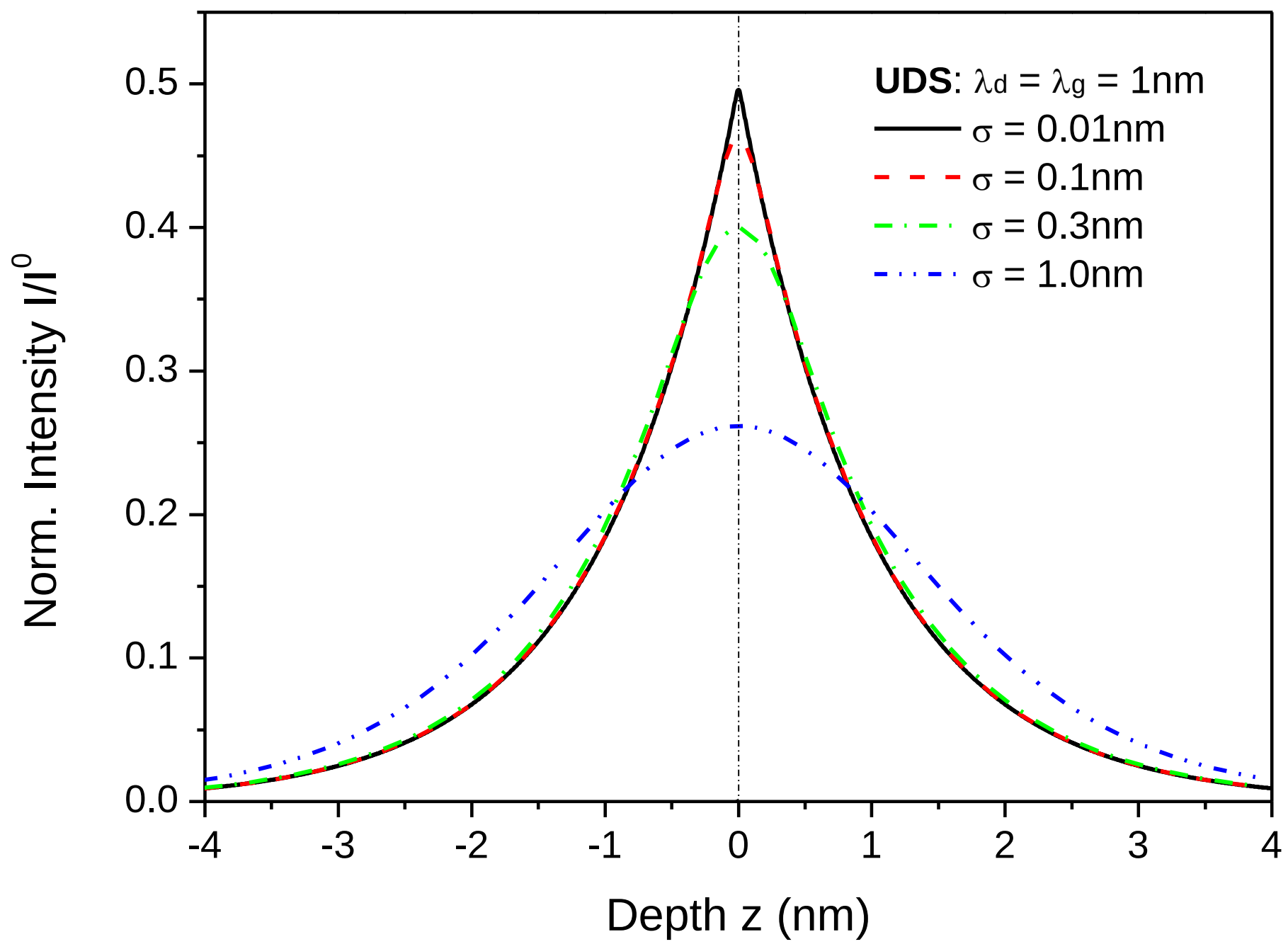

Fig. 3a

(b)

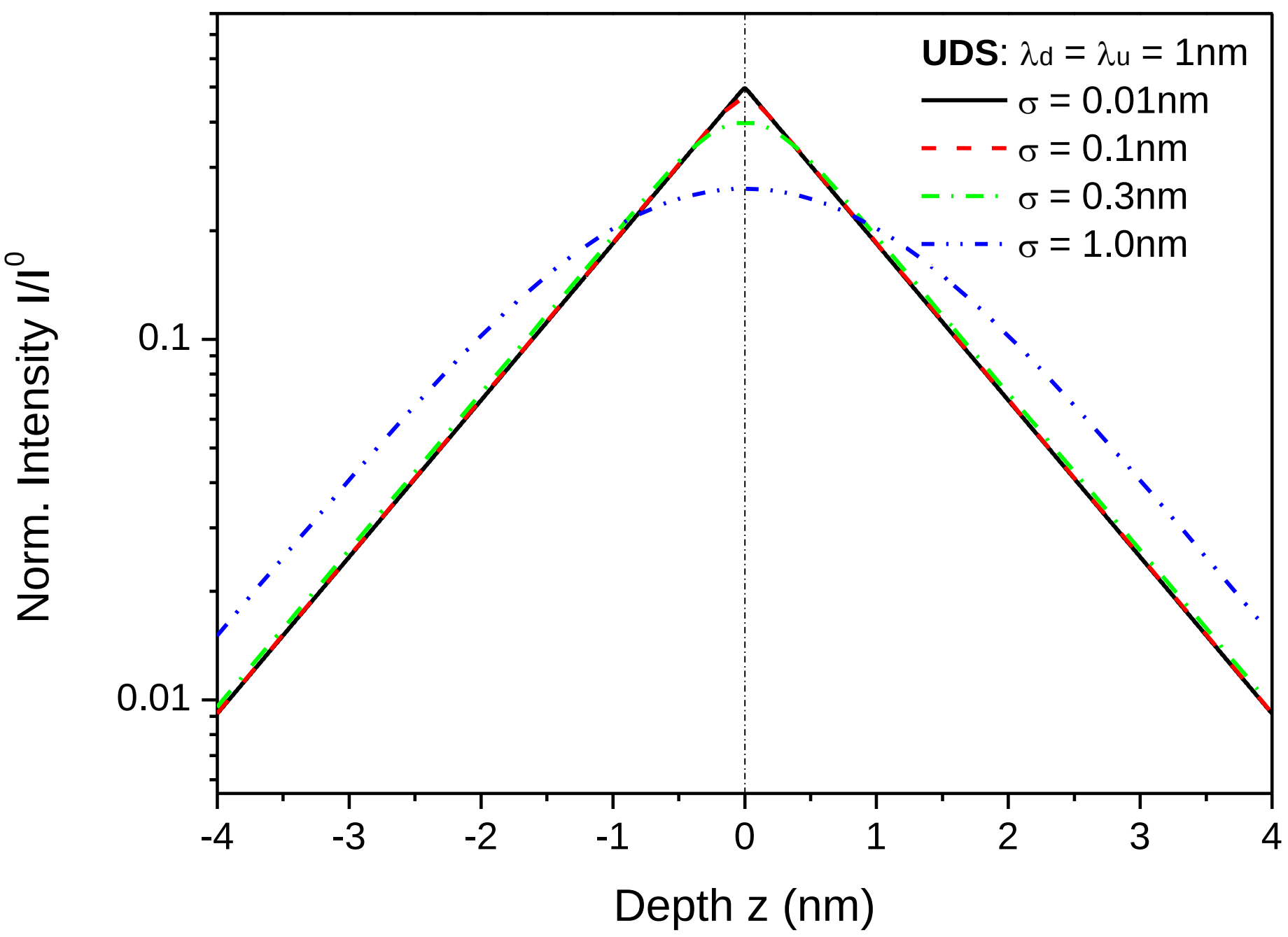

Fig. 3b

(a)

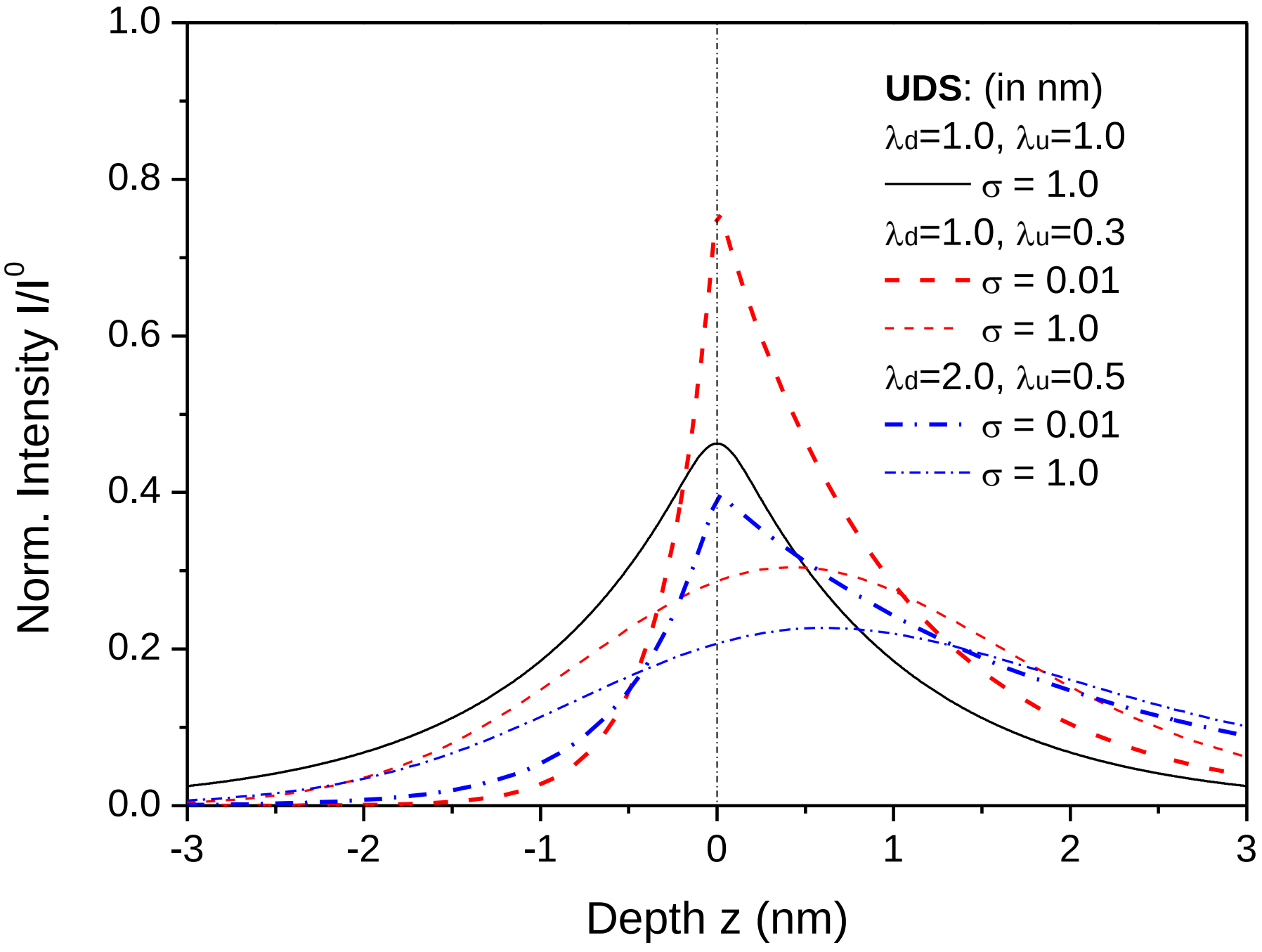

Fig. 4a

(b)

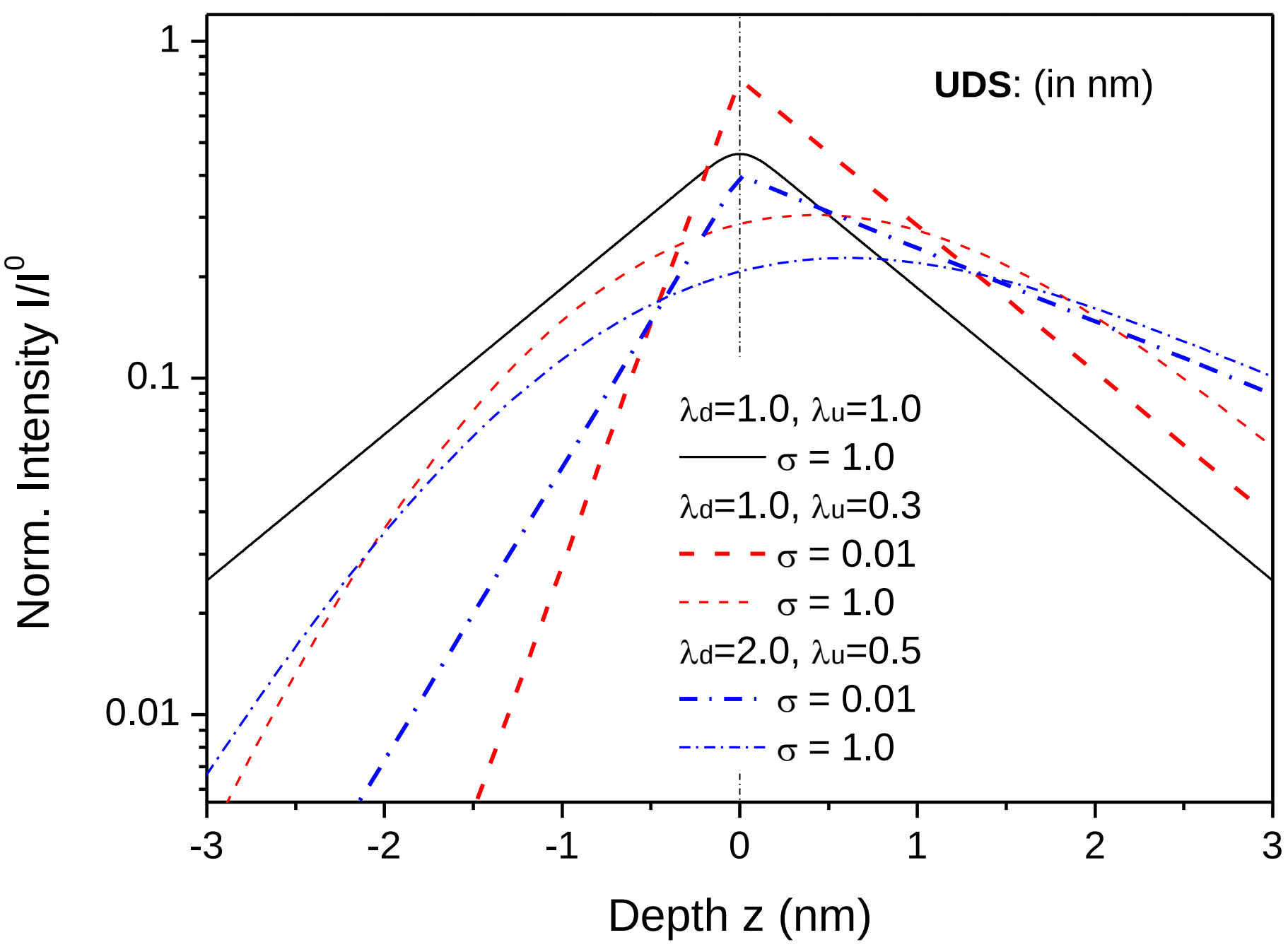

Fig. 4b

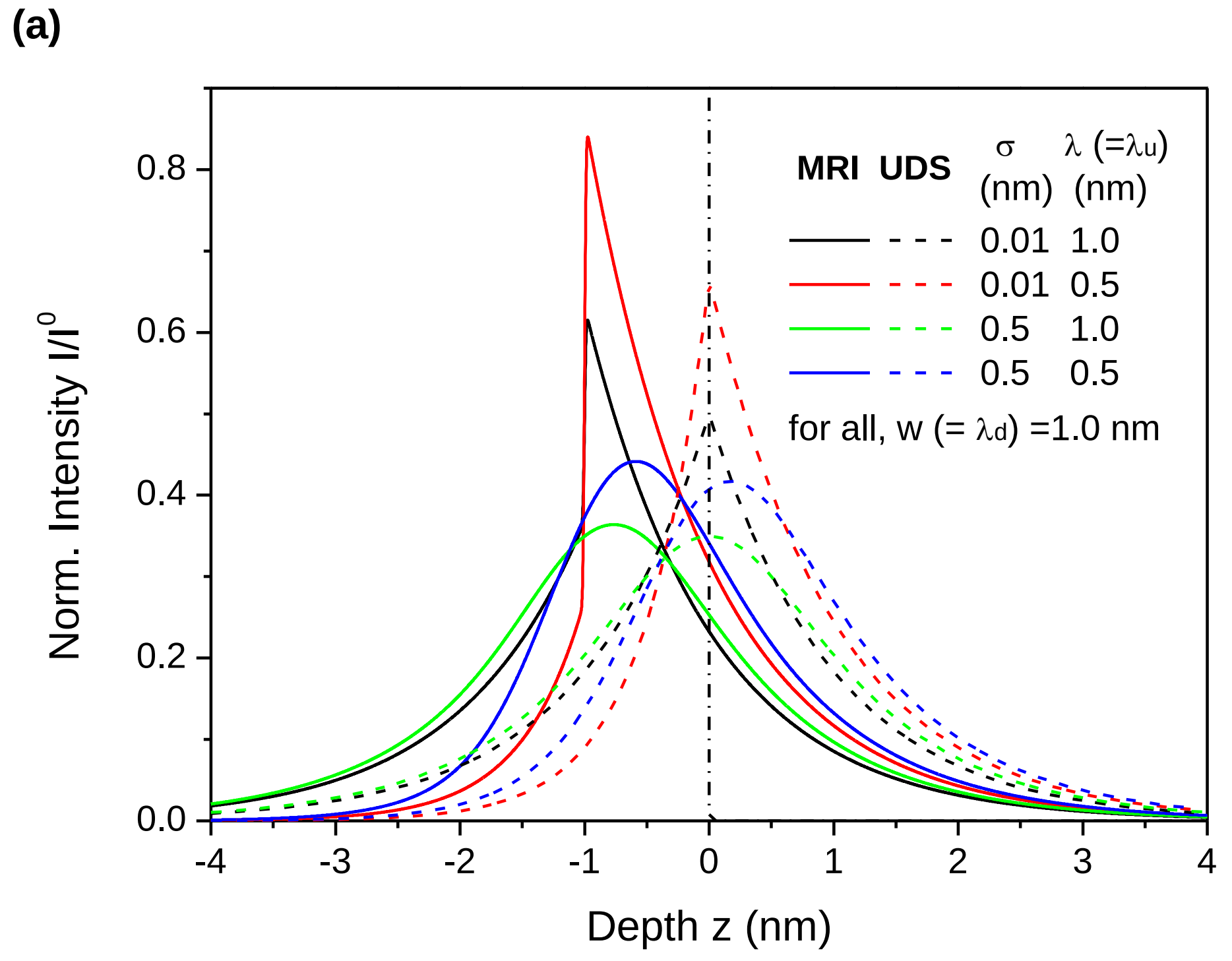

Fig. 5a

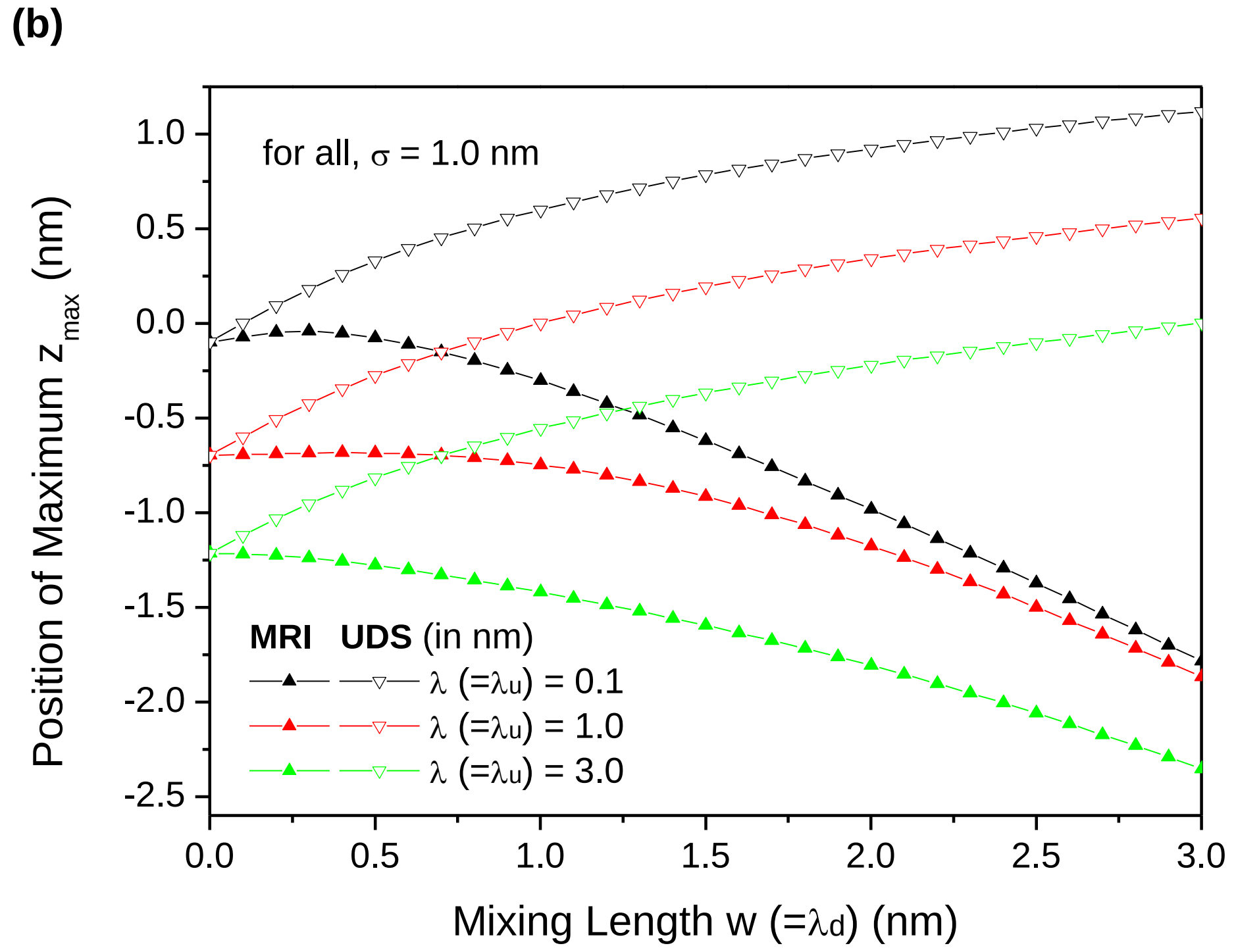
(b)
for all, σ = 1.0 nm
MRI UDS (in nm)
λ (=λu) = 0.1
λ (=λu) = 1.0
λ (=λu) = 3.0
Position of Maximum zmax (nm)
Mixing Length w (=λd) (nm)
1.0
0.5
0.0
-0.5
-1.0
-1.5
-2.0
-2.5
0.0
0.5
1.0
1.5
2.0
2.5
3.0

Fig. 5b

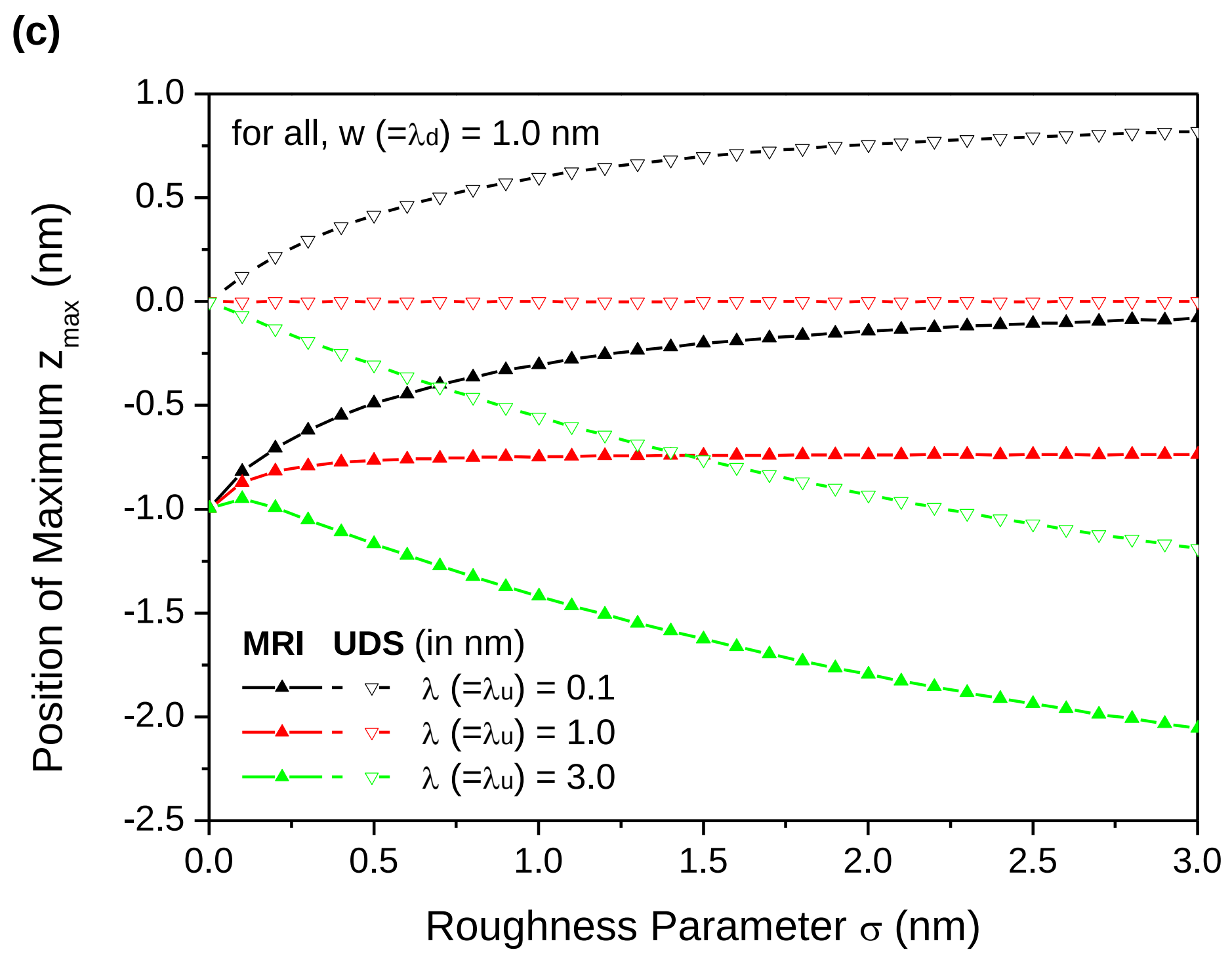
(c)
for all, w (=λd) = 1.0 nm
Position of Maximum z max (nm)
MRI UDS (in nm)
λ (=λu) = 0.1
λ (=λu) = 1.0
λ (=λu) = 3.0
Roughness Parameter σ (nm)

Fig. 5c

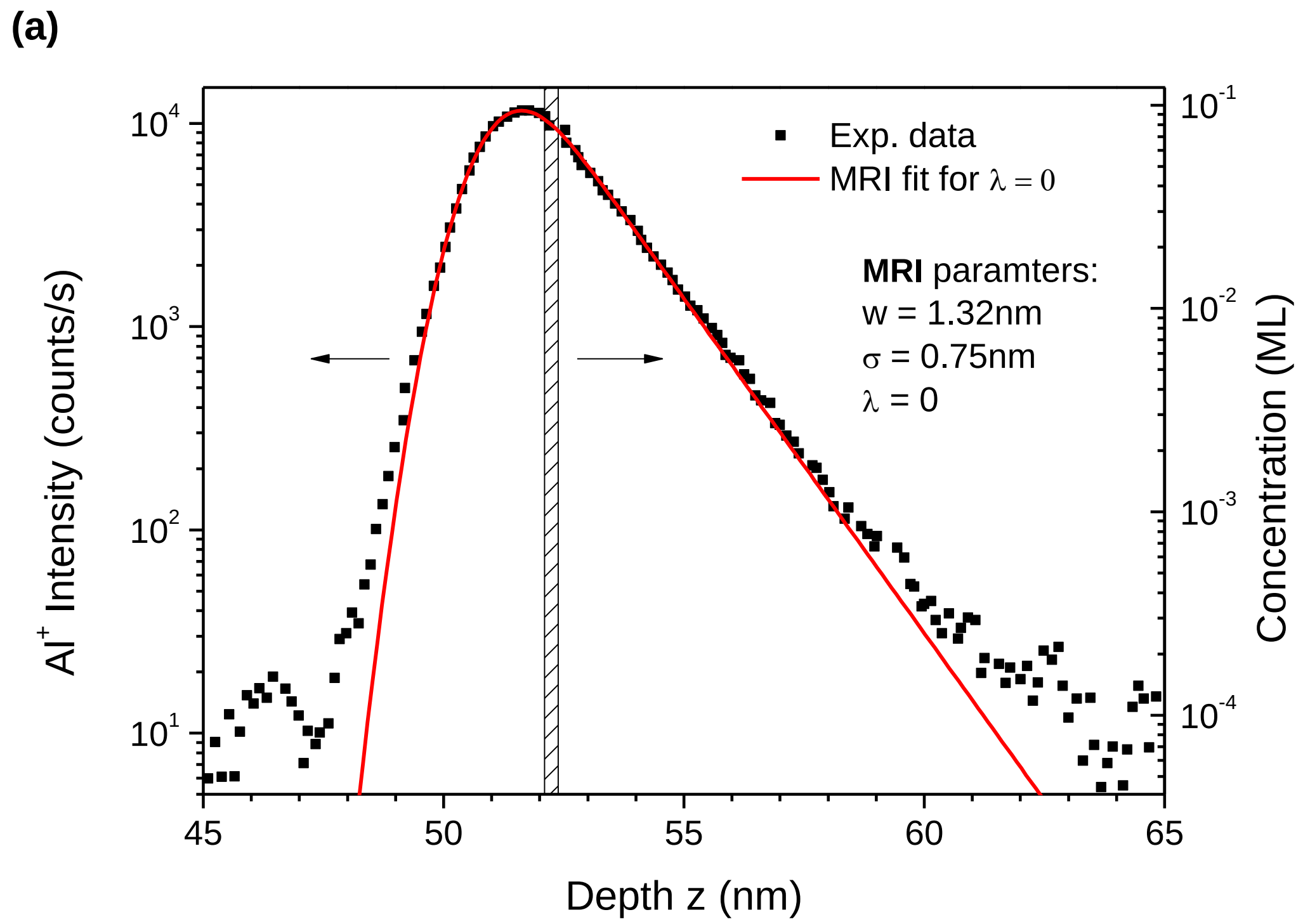

Fig. 6a

**(b)**

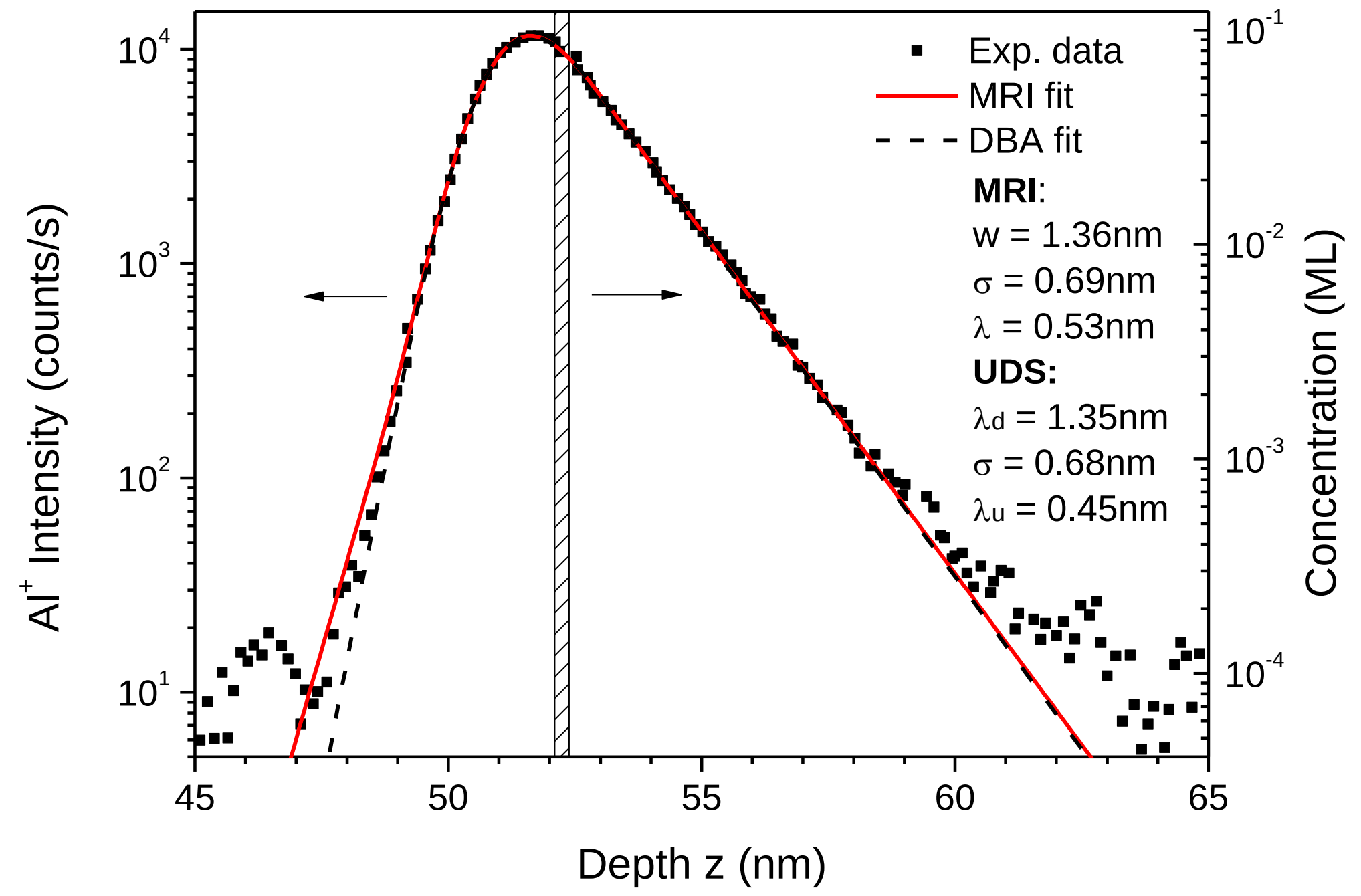

Fig. 6b

**(c)**

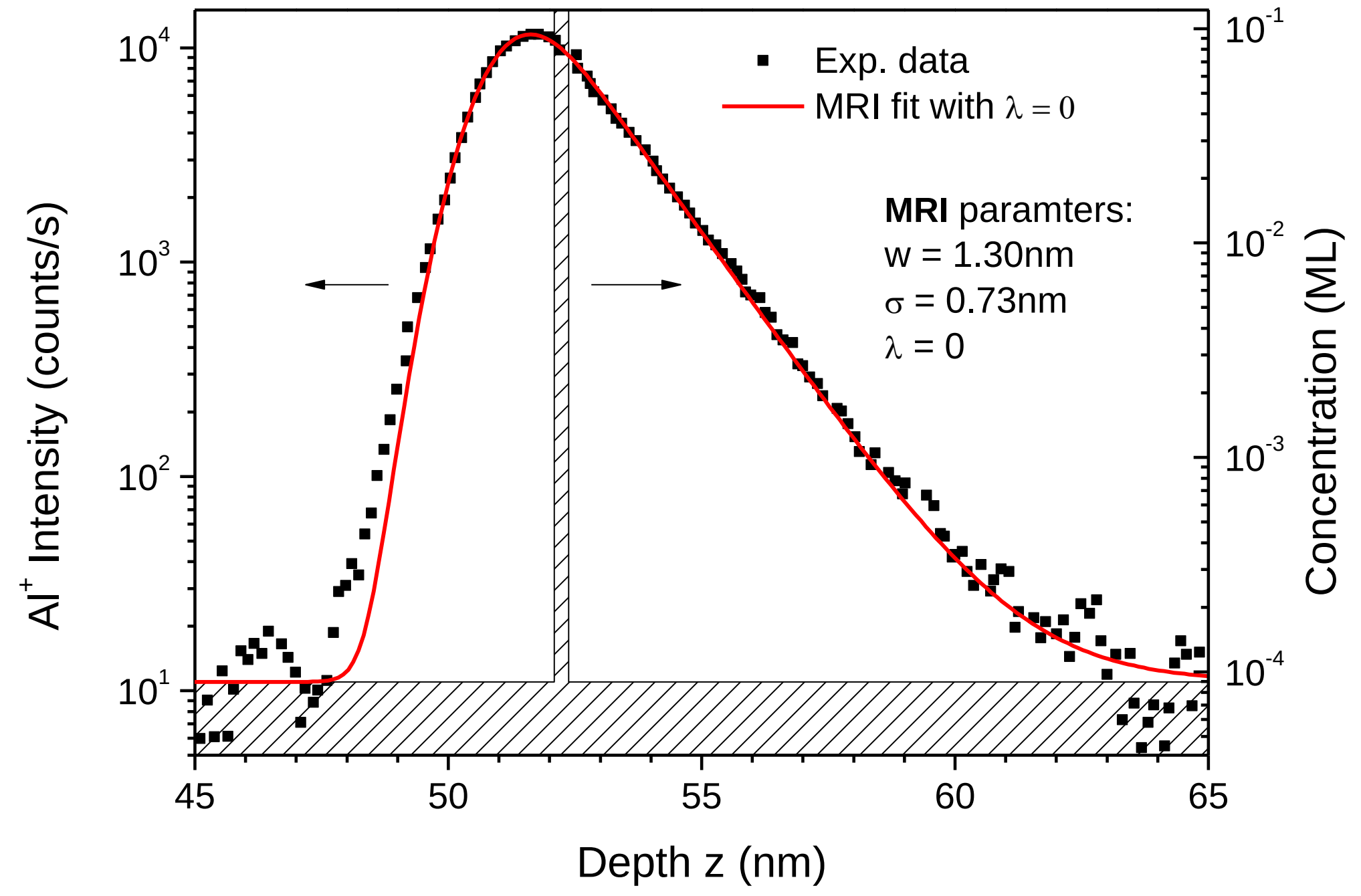

Fig. 6c

**(d)**

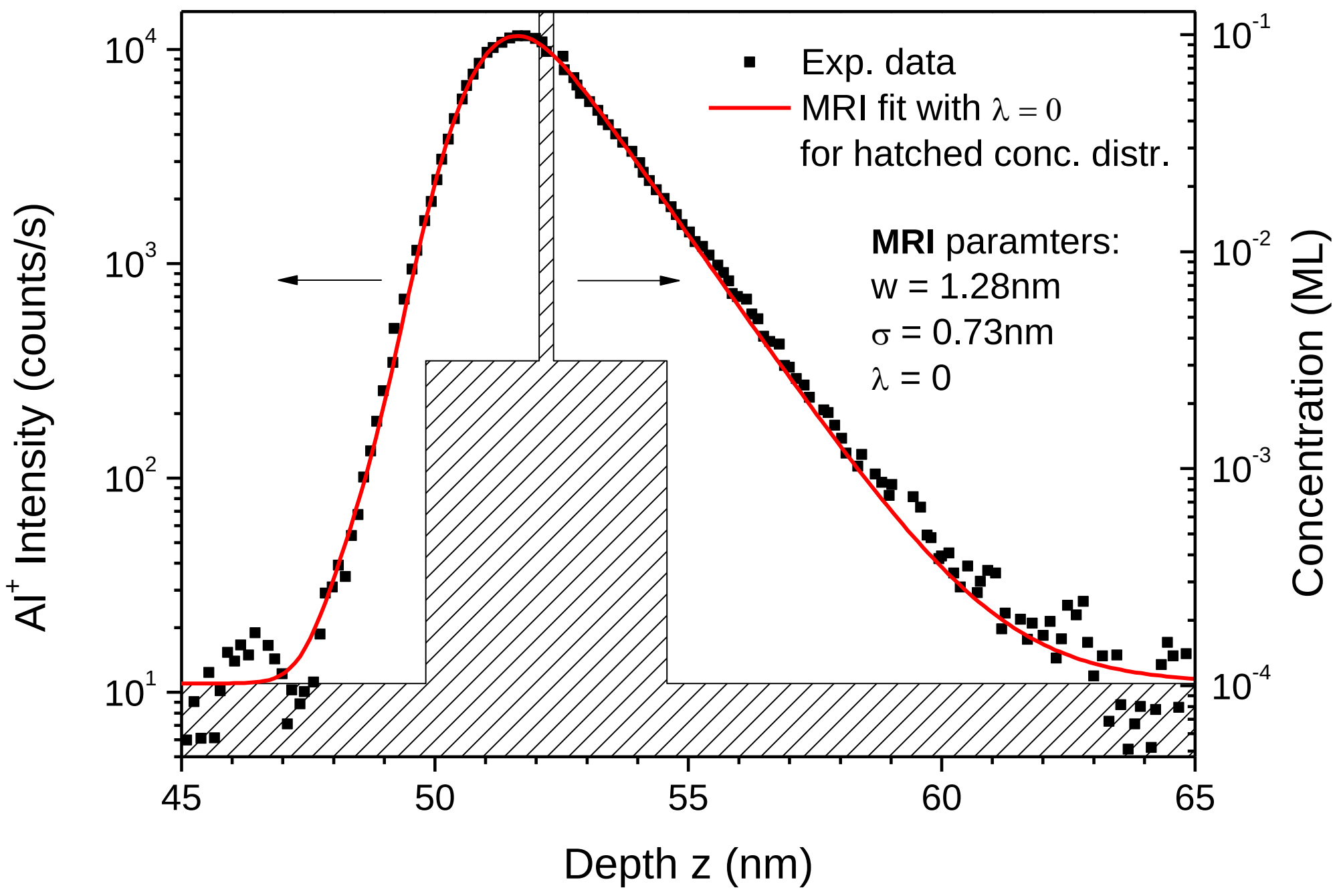

Fig. 6d

**(e)**

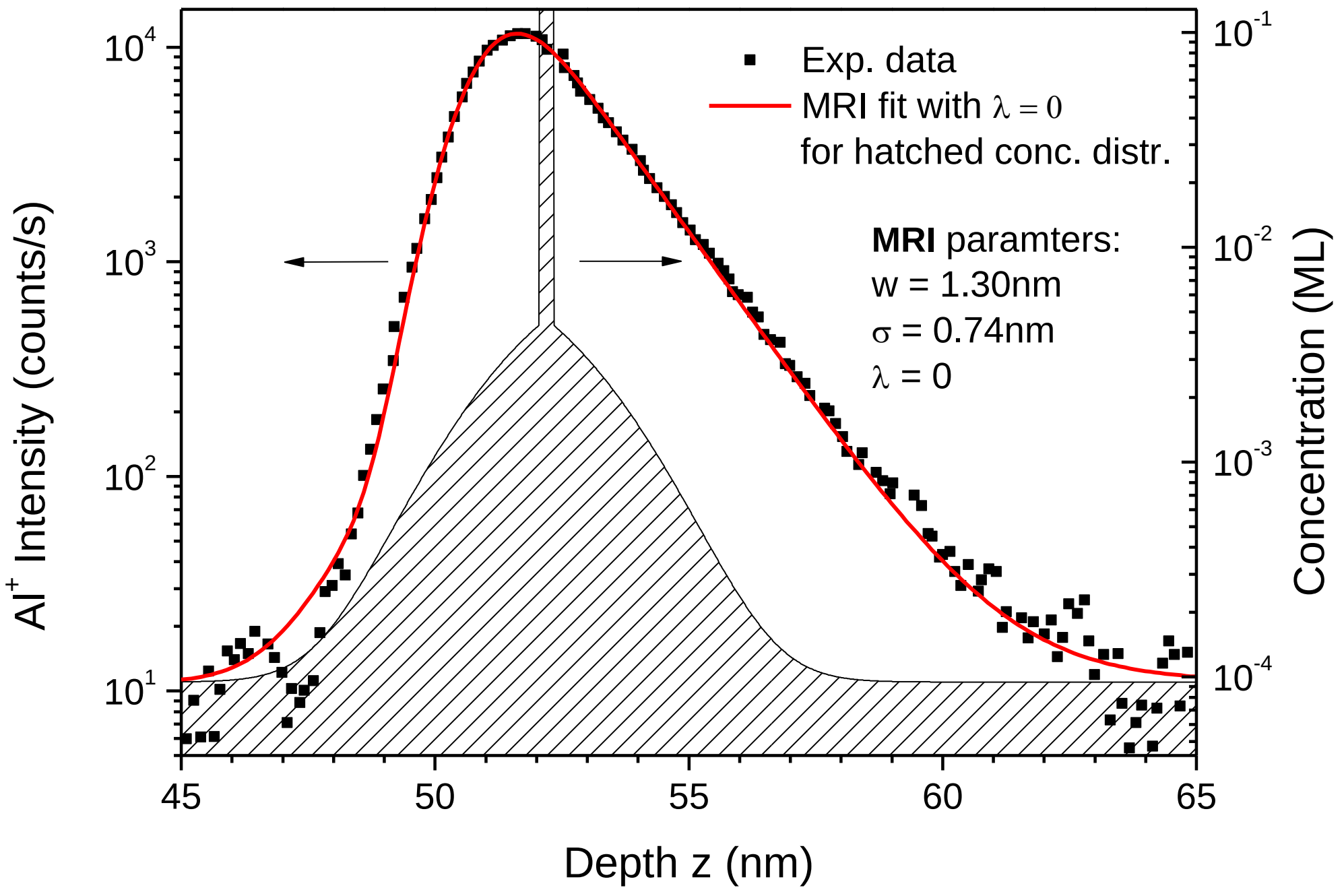

Fig. 6e

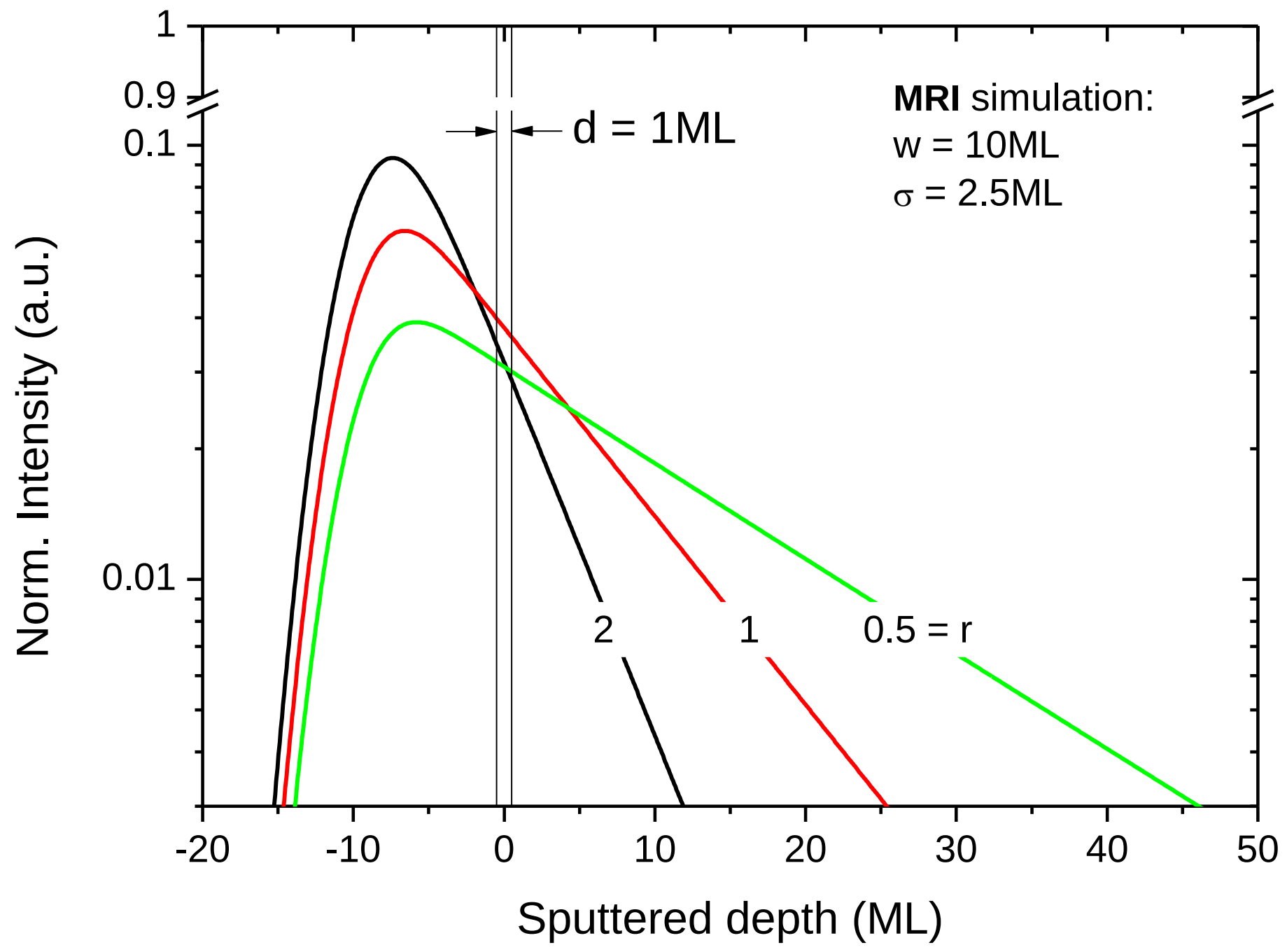

Fig. 7

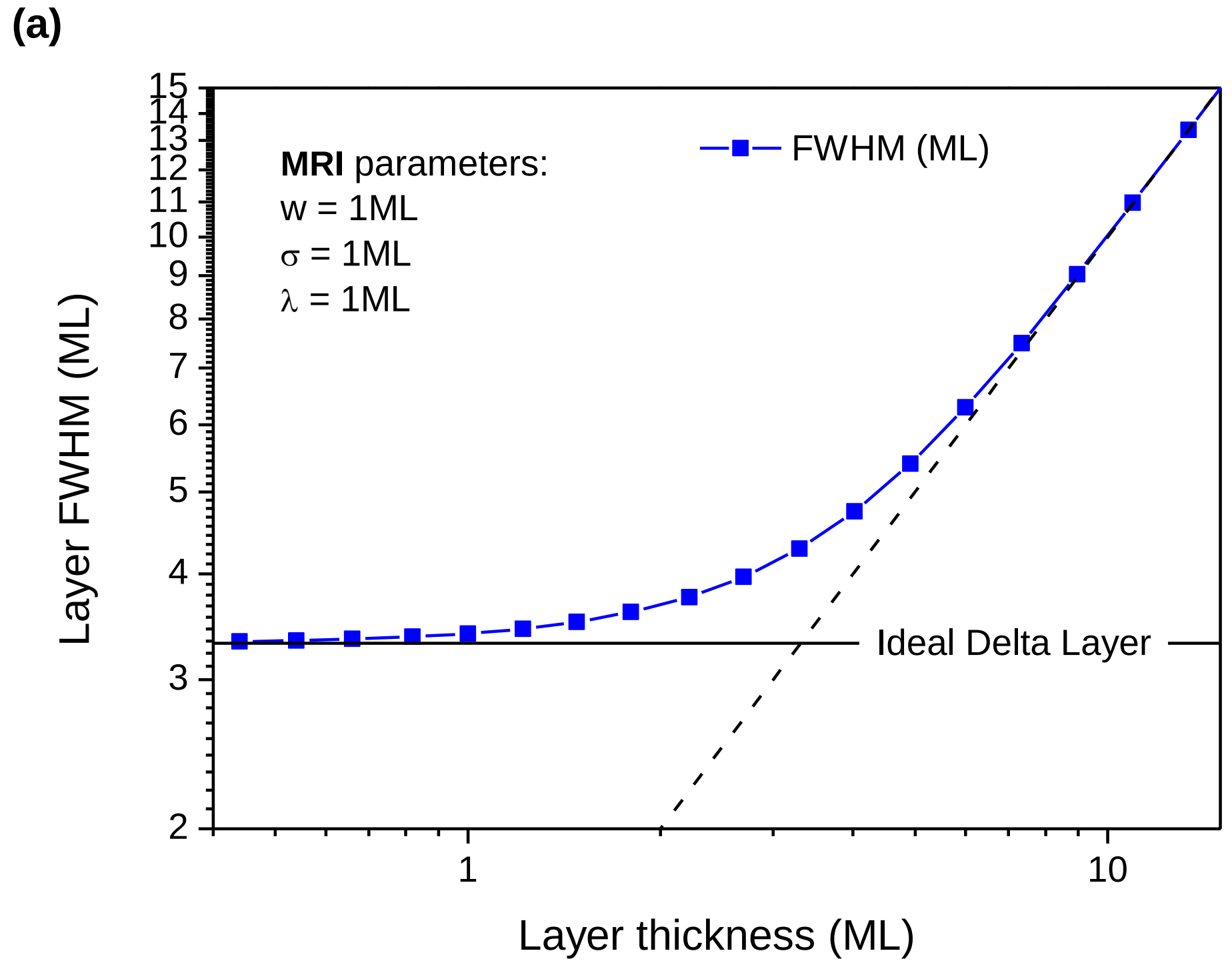

Fig. 8a

**(b)**

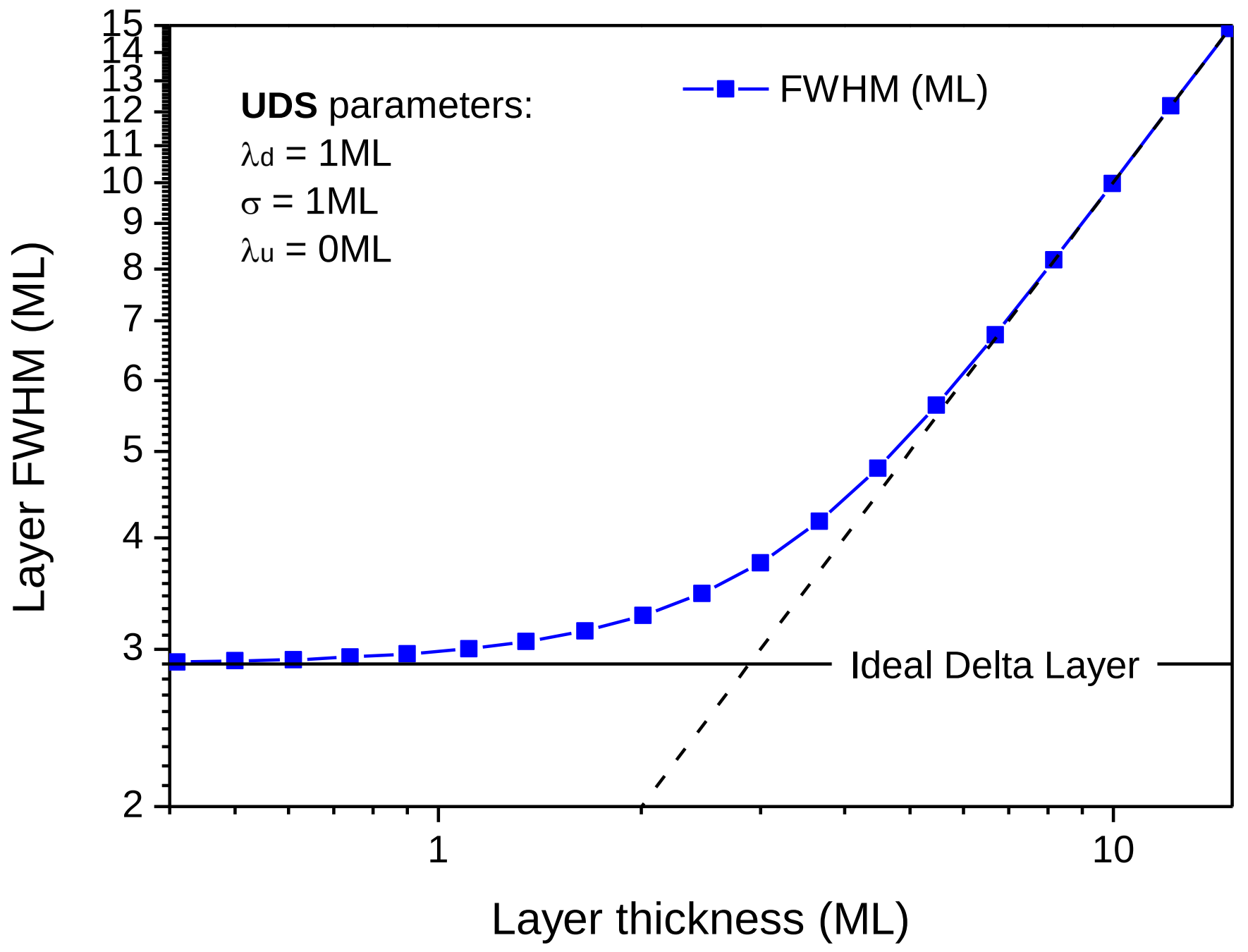

Fig. 8b

**(a)**

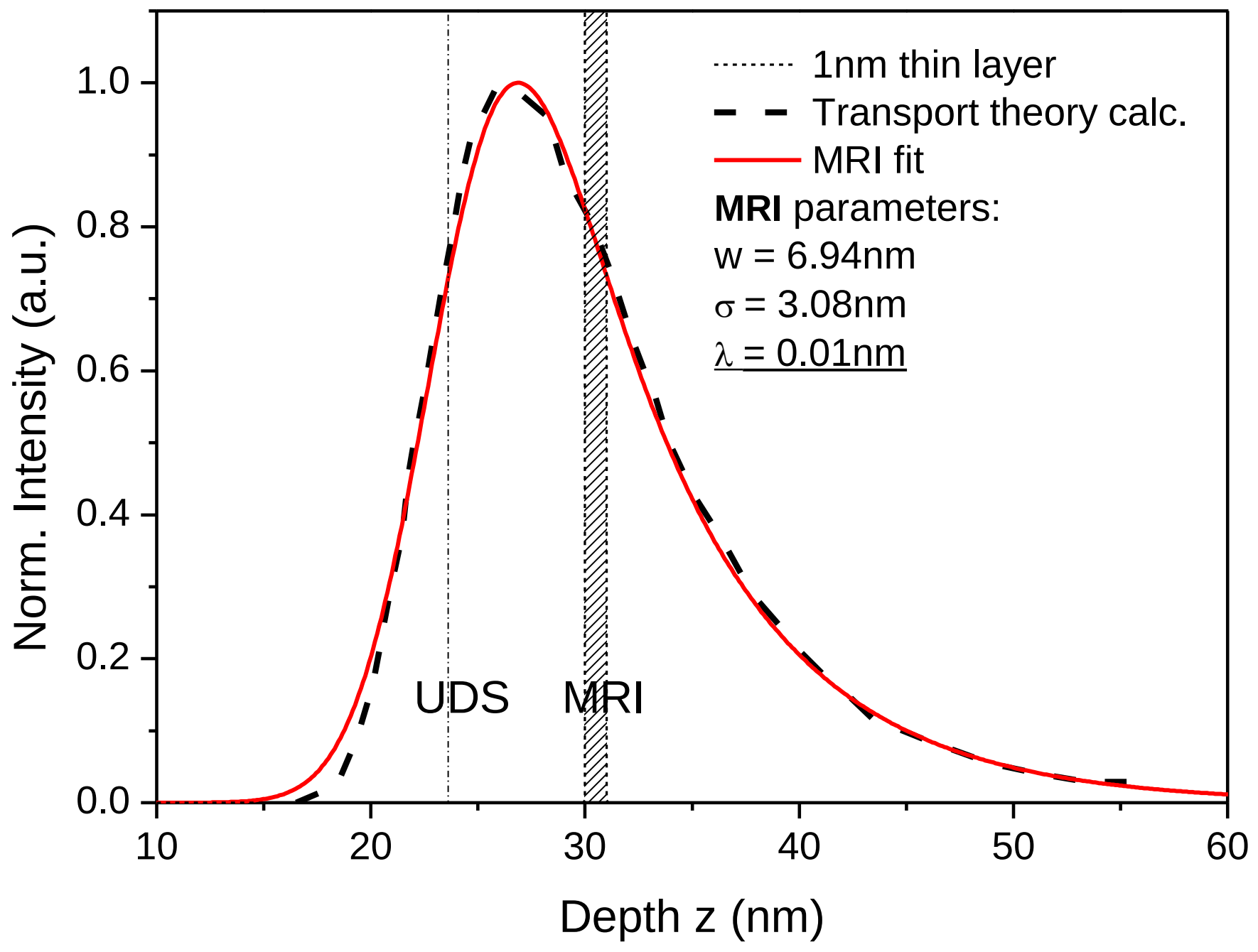

Fig. 9a

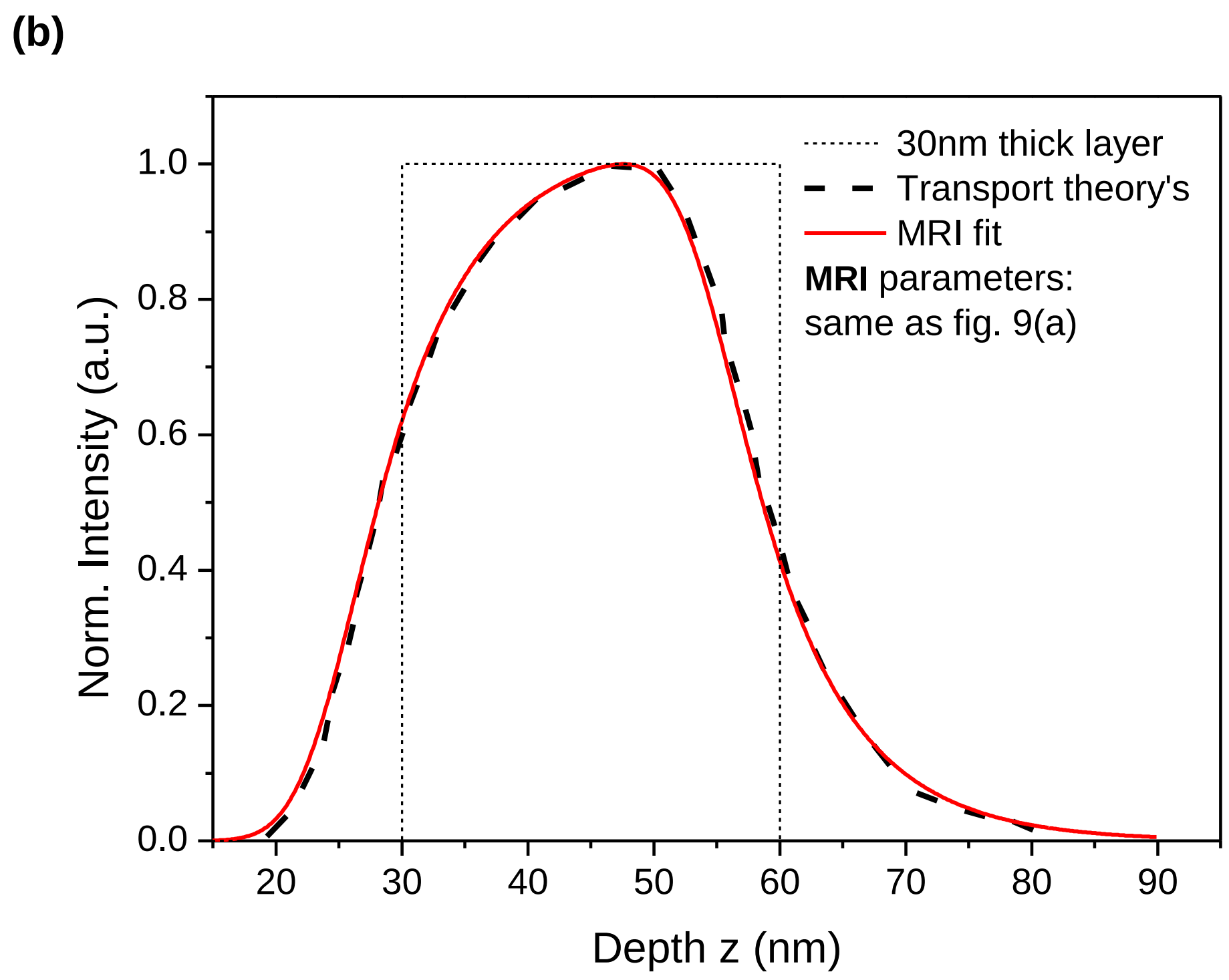

Fig. 9b

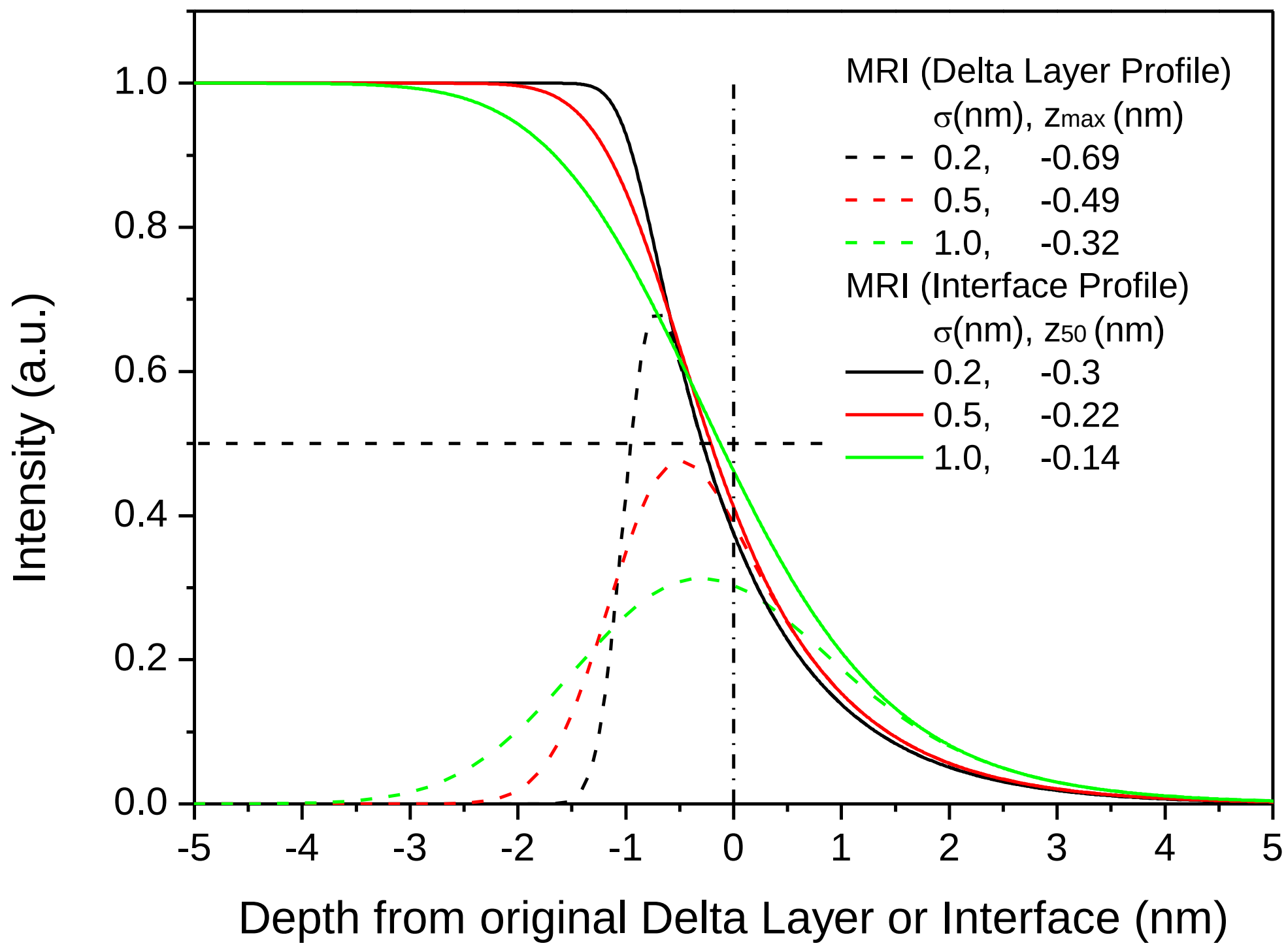
MRI (Delta Layer Profile)
σ(nm), zmax (nm)
0.2, -0.69
0.5, -0.49
1.0, -0.32
MRI (Interface Profile)
σ(nm), z50 (nm)
0.2, -0.3
0.5, -0.22
1.0, -0.14
Intensity (a.u.)
Depth from original Delta Layer or Interface (nm)
1.0
0.8
0.6
0.4
0.2
0.0
-5
-4
-3
-2
-1
0
1
2
3
4
5

Fig. 10

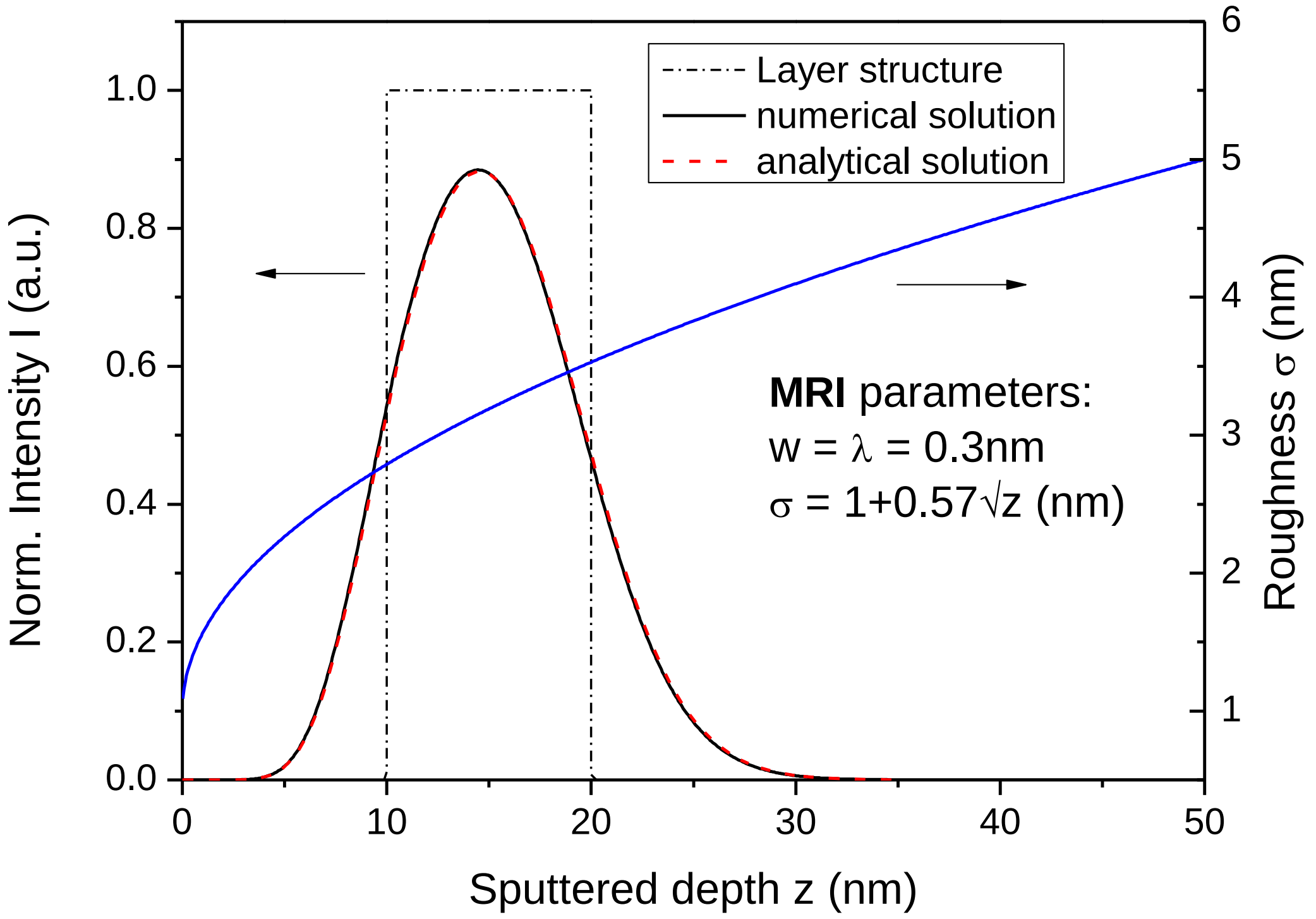

Fig A6.5-1

**Tables and Table captions:**

Table 1:
SIMS depth profile of Al of an AlAs monolayer in GaAs (from Vajo et al. [49]), fitted by the analytical solution of MRI (Fig. 6, a and b, and UDS in Fig. 6b Eqs. (8) and (9), respectively, with layer thickness $d$= 0.28 nm), and by the numerical solution of MRI (Fig. 6, d,e). The respective model parameters for the best fit and the average deviation of the fit (ε(%) (see Appendix 6.4) are presented.

Table 1

| Figure | $w$ (nm) | $\sigma$ (nm) | $\lambda$ (nm) | $\varepsilon$ (%) |
|---|---|---|---|---|
| 6a | 1.32 | 0.75 | 0 | 1.15 |
| 6b (MRI) | 1.36 | 0.69 | 0.53 | 1.10 |
| 6b (UDS) | 1.35 ($\lambda_d$) | 0.68 | 0.45 ($\lambda_u$) | 1.05 |
| 6c | 1.31 | 0.75 | 0 | 1.04 |
| 6d | 1.28 | 0.73 | 0 | 1.02 |
| 6e | 1.30 | 0.74 | 0 | 1.01 |